\documentclass[preprint,12pt]{elsarticle}

\usepackage[outdir=./plots/epstopdf/]{epstopdf}
\usepackage{bm}
\usepackage{amsmath}
\usepackage{graphicx}
\usepackage{subcaption}
\usepackage{multirow} 
\usepackage{makecell} 
\usepackage{pdflscape}
\usepackage{tabularx} 
\usepackage{overpic}
\usepackage{placeins}
\usepackage{caption}
\usepackage[ruled,vlined]{algorithm2e}
\usepackage[justification=centering]{caption}
\usepackage{algpseudocode}
\usepackage{xcolor}

\SetCommentSty{mycommfont}

\usepackage{amssymb}
\usepackage{amsthm}

\usepackage{lineno}

\newtheorem{remark}{Remark}[section]

\theoremstyle{definition}

\biboptions{sort&compress}

\journal{To be determined}

\begin{document}

\begin{frontmatter}



\title{A novel stratified sampler with unbalanced refinement for network reliability assessment}
\author{Jianpeng Chan}
\author{Iason Papaioannou}
\author{Daniel Straub}
\address{Engineering Risk Analysis Group, Technische Universit{\"a}t M{\"u}nchen, Munich, Germany}

\begin{abstract}
%% Text of the abstract 
We investigate stratified sampling in the context of network reliability assessment. 
We propose an unbalanced stratum refinement procedure, which operates on a partition of network components into clusters and the number of failed components within each cluster. 
The size of each refined stratum and the associated conditional failure probability, collectively termed failure signatures, can be calculated and estimated using the conditional Bernoulli model. 
The estimator is further improved by determining the minimum number of component failure $i^*$ to reach system failure and then consider only strata with at least $i^*$ failed components. 
We propose a heuristic but practicable approximation of the optimal sample size for all strata, assuming a coherent network performance function. 
The efficiency of the proposed stratified sampler with unbalanced refinement (SSuR) is demonstrated through two network reliability problems. 
\end{abstract}

\end{frontmatter}

\graphicspath{{figures/}{tikz/}{plots/}{epstopdf/}} 


\section{Introduction}
% A review of the stratified sampling for network reliability analysis 
A primary goal of network reliability assessment is to estimate the probability $p_f$ that a network fails to meet the required performance under disturbances. 
This probability estimate is central to reliability-based network design and for quantifying a network's resilience \cite{Li&Liu2021}. 

The network failure $F$ is often assessed by a network performance function $g(\bm{x})$ and a specified threshold $\gamma$. 
Herein, the vector $\bm{x} = \{x_1,\cdots,x_n\}$ denotes the states of $n$ network components. 
The meaning of $g(\bm{x})$ and $\gamma$ vary over different settings. 
% For instance, when analyzing the source-target connectivity in a water supply system, $g(\bm{x})$ is a binary indicator function equal to 0 when the target node is disconnected from the source and equal to 1 otherwise. 
% In this context, $\bm{x}$ is the binary damage states of pipelines. 
% The network fails when $g(\bm{x}) \leq 0$. 
For instance, in the probabilistic contingency analysis of a power grid, $g(\bm{x})$ can compute the percentage blackout size of the grid \cite{Billinton&Li1994}, and the threshold $\gamma$ is chosen according to the regulation or the requirements of the operator or regulator. 
The network failure occurs when the percentage blackout size exceeds the threshold $\gamma$. 
Due to the uncertainty embedded in both material and external disturbances, the component states are not deterministic. 
Hence, it is more appropriate to model the components' state as a random vector $\bm{X} = \{X_1, \cdots, X_n\}$. 

We denote the probability mass function and the sample space of $\bm{X}$ as $p_{\bm{X}}(\bm{x})$ and $\Omega_{\bm{X}}$, respectively. 
Consequently, the failure probability can be written as follows: 
\begin{equation}
    p_f = \Pr(F) = \sum_{\bm{x} \in \Omega_{\bm{X}}} \mathbb{I}\{ \bm{x} \in F \} p_{\bm{X}}(\bm{x}),
\end{equation} 
where the indicator function $\mathbb{I}\{ \bm{x} \in F\}$ equals one if $\bm{x}$ leads to system failure and zero otherwise. 
Despite its simple form, estimating $p_f$ is a challenging task. 
Even for elementary network performance functions (or metrics), such as connectivity and maximum flow, the exact calculation of $p_f$ is NP-hard \cite{Provan&Ball1984, Ball1986} for a general network. 
For physics-driven network performance, the computation becomes even more challenging since these performance functions are often costly to evaluate. 
One has to resort to practically efficient enumeration or approximation methods that deliver either probability bounds or a sample estimate of $p_f$. 
Overall, no single method dominates in all scenarios so the choice should be based on the specific problem at hand. 

Examples of practically efficient algorithms include cut(or path)-based methods \cite{Jane&others1993, Zuo&others2007, Brown&others2021}, binary decision diagrams \cite{Imai&others1999, Hardy&others2007}, universal generating function methods \cite{Levitin&others2003, Li&Zio2012}, matrix-based methods \cite{Song&Kang2009}. 
These methods converge to the true value of $p_f$. 
However, they are in general not suitable for high-dimensional problems, and their efficiency depends on the specific characteristics of the network performance function. 
For instance, in the context of binary decision diagrams, the problem should allow for efficient construction of the diagram and the selection of an appropriate variable ordering \cite{Hardy&others2007}. 
% In the context of cut-based methods, the performance function must be coherent and allow for an efficient search through all minimal cuts. 
% In the context of the universal generating function methods, the efficiency arises from the hierarchical structure of the network performance function, which facilitates the recursive combination of like terms. 
% Lastly, the matrix-based method also relies on the explicit expression of the failure event as unions, intersections, and complements of the component event.

By contrast, approximation methods are more general and applicable to higher dimensions, but they only provide approximate results, either in the form of probability bounds or an estimator of $p_f$. 
In principle, cut(or path)-based methods can be terminated prematurely or work with incomplete minimal cuts to yield a bound of $p_f$. 
Alternatively, the recursive decomposition method \cite{Li&He2002, Lim&Song2012, Paredes&others2018, Byun&others2024} refines the lower and upper bound of $p_f$ by iteratively separating the survival and failure domain from the current unspecified domain. 
% This is also known as the state space decomposition method in operations research. 
While achieving good results in low to moderate dimensional problems, the convergence rate of the bounds, especially the upper bound of $p_f$, degenerates in high dimensional problems. 
In addition, the recursive decomposition method can only be applied to coherent performance functions. 

In cases where the specific structure or properties of a network performance function are unclear, sampling-based methods appear promising. 
These include crude Monte Carlo simulation (MCS) \cite{Billinton&Li1994}, subset simulation \cite{Zio&Pedroni2008, Zuev&others2015, Jensen&Jerez2018, Chan&others2022a}, cross-entropy-based importance sampling \cite{Hui&others2005, Chan&others2023a, Chan&others2024}, stratified sampling \cite{VanSlyke&Frank1971}. 
The resulting estimators are often consistent, meaning they converge to the true $p_f$ with an increasing sample size. 
However, when the sample size is small, these estimators can be skewed and have large variance, so many samples may be needed to achieve an accurate result \cite{Chan&others2025}. 
We note that for connectivity or maximum-flow-based problems, other efficient sampling methods exist, such as the counting-based algorithm \cite{Duenas&others2017}, creation-process-based methods \cite{Elperin&others1991, Cancela&others2022}, or recursive variance reduction \cite{Cancela&Khadiri1995}. 
Meanwhile, actively trained surrogate models \cite{Dehghani&others2021, Ding&others2024} and sampling-based signature methods \cite{Coolen&Coolen-Maturi2024, DiMaio&others2023} also gain increasing attention in network reliability assessment. 
% While the surrogate model may predict the network performance prediction poorly, it can still be used to identify critical samples for running the expensive network model. 

This paper investigates stratified sampling in the context of network reliability assessment. 
In particular, we consider a general network performance function with independent binary inputs, where $x_i = 1$ denotes the failure of the $i$-th component and $x_i = 0$, otherwise. 
% In practice, components can have multiple failure states and can fail dependently. 
% Extending the proposed method to account for these challenges will be reserved for future work. 
Stratification is a well-known variance reduction technique that has proven successful in many fields, including structural reliability assessment, survey sampling, and other areas of applied mathematics \cite{Tong2006, Shields&others2015, Cochran1977, Song&Kawai2023, Etore&others2011, Pettersson&Krumscheid2022}. 
The use of stratified sampling for addressing network reliability problems has also been investigated \cite{VanSlyke&Frank1971, Fishman1989, Hui&others2003}. 
Stratified sampling is also connected to the system signature and, as we show, the proposed method can be used to determine this signature. 
The main novel contribution of our work, however, is as follows: 
(1) For independent binary components, whether identical or not, we introduce a novel strata refinement strategy. 
The proposed strategy is based on the number of failed components within different clusters of a progressively refined partition. 
% While this approach is related to the signature-based method \cite{Patelli&others2017, DiMaio&others2023}, it does not require components in each group to be of the same type or exchangeable. 
% The failure probability of each component can be unique.   
We further employ the conditional Bernoulli model to sample conditional on each stratum and to calculate the size (or probability volume) of each stratum. 
This contribution is detailed in Subsections \ref{Subsubsec: Independent but non-identical components} and \ref{Subsec: one-step refinement procudure}. 
(2) We introduce an approximation strategy of the unknown optimal sample size in each stratum in Subsection \ref{Subsubsec: Approximating the optimal sample allocation}. 
(3) For physics-based performance functions, we use a genetic algorithm to find the minimum number of failed components required to cause the system failure, denoted as $i^*$. 
States with less than $i^*$ failed components can then be safely removed, often resulting in a significant improvement of the algorithm efficiency. 
This approach is presented in Section \ref{Subsec: Stratified sampler after excluding the redundant strata}. 
In addition, we demonstrate that, under proportional or optimal sample allocation, refining any stratum in stratified sampling does not increase the variance. 
This ensures that the variance ratio relative to conditional Monte Carlo remains non-increasing during refinement. 
This property was also observed by Pettersson and Krumscheid \cite{Pettersson&Krumscheid2022}.  
% However, their proof appears too concise to be easily understood.
To enhance clarity, we present an independent and more detailed proof in this work, from which the necessary and sufficient conditions for the variance ratio to strictly decrease can be derived. 

The paper is organized as follows: Section 2 introduces the basic ideas and implementations of the stratified sampler. 
Sections 3 and 4 contribute two improvements: the removal of redundant strata and stratum refinement. 
In particular, we introduce methods for identifying $i^*$ in both connectivity- and physics-based problems and detail the stratum refinement procedure. 
A summary and a workflow of the stratified sampler are also included in Section 4. 
Finally, the efficiency of the proposed stratified sampler is investigated in Section 5 through two numerical examples: one related to power flow analysis and the other to a water supply system. 

\section{Stratified sampling for network reliability}
\label{Sec: classic stratified sampler}
We first present the basic idea of the standard stratified sampler, with a focus on its application in network reliability assessment. 
This is followed by implementation details, including the conditional Bernoulli model, randomization of the fractional sample size, and the heuristic for approximating optimal sample allocation. 
% While these details can be implemented with the two improvements discussed in Sections 3 and 4, they are introduced here for conciseness. 
\subsection{Stratified sampling estimator}
\label{Subsec: analysis of the classic stratified sampler}
According to the total probability theorem, it holds that: 
\begin{equation}
\label{Eq: Total probability theorem}
    p_F = \Pr(F) = \sum_{i=0}^n \Pr(I=i) \Pr(F \mid I=i),
\end{equation}
where $I$ is a random variable denoting the number of failed components. 
% It is evident that the system should not fail when all components are functional, i.e., when $I=0$. 
For conciseness, we introduce the following notation:
\begin{align}
\lambda_i &\triangleq \Pr(I=i), \\
p_{F \mid i} &\triangleq \Pr(F \mid I=i).
\end{align}
In many cases, the probability of having $i$ failed components, $\lambda_i$, can be calculated accurately in advance, so the problem of estimating the failure probability $p_F$ becomes equivalent to estimating a set of conditional probabilities, $p_{F \mid i=0},\cdots, p_{F \mid i=n}$. 
% The rationale behind such decomposition is that the conditional probabilities are often significantly larger than $p_F$ and, hence, are easier to estimate using sampling-based methods. (not necessarily true) 
The standard stratified sampling estimator is obtained when the conditional probabilities are estimated by crude MCS, where each stratum is characterized by a specified number of failed components. 
In this context, $\lambda_i>0$ is the size (or probability volume) of the $i$-th stratum. 
This stratified estimator, denoted as $\widehat{p}_F^{(\text{SS})}$, can be expressed as 
\begin{equation}
\label{Eq: Classic stratified sampling}
    \widehat{p}_F^{(\text{SS})} = \sum_{i=0}^n \lambda_i \widehat{p}_{F\mid i} = \sum_{i=0}^n \frac{\lambda_i}{N_i} \sum_{k=1}^{N_i} \mathbb{I}\{ \bm{x}_k^{(i)} \in F\}, \quad \bm{x}_k^{(i)} \sim p_{\bm{X} \mid i}(\bm{x}). 
\end{equation}
Here, $N_i$ denotes the number of samples allocated to the $i$-th stratum for estimating the conditional probability $p_{F\mid i}$. 
The resulting crude MCS estimator is denoted as $\widehat{p}_{F\mid i}$. 
Moreover, $p_{\bm{X}\mid i}(\bm{x}) \propto p_{\bm{X}}(\bm{x}) \mathbb{I}\{I=i\}$ is the input distribution of $\bm{X}$ given there are $i$ failed components. 
The variance of $\widehat{p}_F^{(\text{SS})}$ can be computed by
\begin{equation}
\label{Eq: The variance of the classic stratified sampling estimator}
    \mathbb{V}\left( \widehat{p}_F^{(\text{SS})} \right) 
    = \sum_{i=0}^n \frac{\lambda_i^2}{N_i} p_{F\mid i}(1-p_{F\mid i}) 
    = \sum_{i=0}^n \frac{\lambda_i^2}{N_i} p_{F\mid i}- \sum_{i=0}^n \frac{\lambda_i^2}{N_i}p^2_{F\mid i}.
\end{equation}

\subsection{Allocation of samples} 
When the sample size for each stratum is set in proportion to the probability of that stratum, i.e., 
\begin{equation}
\label{Eq: The proportional budget allocation strategy}
N_i = N \lambda_i \triangleq N_i^{(\text{prop})}, \quad i = 0,\cdots, n,
\end{equation}
the variance in Eq.~(\ref{Eq: The variance of the classic stratified sampling estimator}) becomes 
\begin{align}
\label{Eq: The variance of the classic stratified sampling estimator (proportional allocation)}
    \mathbb{V}\left( \widehat{p}_F^{(\text{SS,prop})} \right) 
    & = \frac{1}{N} \sum_{i=0}^n \lambda_i p_{F\mid i} - \frac{1}{N} \sum_{i=0}^n \lambda_i p^2_{F\mid i} \notag \\
    & \leq \frac{1}{N} \sum_{i=0}^n \lambda_i p_{F\mid i} - \frac{1}{N} \left( \sum_{i=0}^n \lambda_i p_{F\mid i} \right)^2 = \mathbb{V}\left( \widehat{p}_F^{(\text{MCS})} \right). 
\end{align} 
Here, $N$ is the overall sample size. 
Inequality~(\ref{Eq: The variance of the classic stratified sampling estimator (proportional allocation)}) implies that the variance of the stratified sampling estimator with proportional sample allocation $\{N_i^{(\text{prop})}\}_{i=0}^n$, denoted as $\widehat{p}_F^{(\text{SS, prop})}$, is not larger than that of crude MCS \cite{Papaioannou2021}. 

The variance of the stratified sampling estimator can be further reduced by employing a different sample allocation strategy. 
In particular, the optimal sample allocation strategy is given by 
\begin{equation}
\label{Eq: The optimal budget allocation strategy}
    N_i = N \frac{ \lambda_i \sqrt{p_{F \mid i} (1-p_{F \mid i})} }{ \sum_{k=0}^{n}\lambda_k \sqrt{p_{F \mid k} (1-p_{F \mid k})} } \triangleq N_i^{(\text{opt})}, \quad i=0,\cdots,n.
\end{equation}
Now, the optimal sample size $N_i^{(\text{opt})}$ is proportional to the probability of the $i$-th stratum $\lambda_i$ multiplied by the local standard deviation $\sqrt{ p_{F\mid i}(1-p_{F\mid i}) }$, i.e., the standard deviation of $\mathbb{I}\{ \bm{X} \in F\}$ with $\bm{X} \sim p_{\bm{X}\mid i}(\bm{x})$. 
The corresponding minimum variance is 
\begin{equation}
\label{Eq: The minimum variance of the classic stratified sampling estimator}
    \mathbb{V}\left( \widehat{p}_F^{(\text{SS,opt})} \right) 
    = \frac{1}{N} \left( \sum_{i=0}^{n}\lambda_i \sqrt{p_{F \mid i} (1-p_{F \mid i})} \right)^2. 
\end{equation}
Here, $\widehat{p}_F^{(\text{SS, opt})}$ denotes the stratified estimator with optimal sample allocation. 

The degree of variance reduction achieved through stratified sampling can be quantitatively assessed by the ratio of the estimator variance compared to that of crude MCS, that is, $\frac{\mathbb{V}\left( \widehat{p}_F^{(\text{SS})} \right)}{\mathbb{V}\left( \widehat{p}_F^{(\text{MCS})} \right)}$ \cite{Fishman1989}. 
The smaller the variance ratio, the more significant the variance reduction. 
In particular, one can find:
\begin{align}
    \label{Eq: Variance ratio 1}
    r_{ \frac{\text{SS,prop}}{\text{MCS}}} 
    & = \frac{\mathbb{V}\left( \widehat{p}_F^{(\text{SS, prop})} \right)}{\mathbb{V}\left( \widehat{p}_F^{(\text{MCS})} \right)}  = \frac{ \sum_{i=0}^n \lambda_i p_{F \mid i} (1-p_{F \mid i}) }{ p_F - p_F^2},\\ 
    \label{Eq: Variance ratio 2}
    r_{ \frac{\text{SS,opt}}{\text{MCS}}} & = \frac{\mathbb{V}\left( \widehat{p}_F^{(\text{SS,opt})} \right)}{\mathbb{V}\left( \widehat{p}_F^{(\text{MCS})} \right)} = \frac{\left( \sum_{i=0}^{n}\lambda_i \sqrt{p_{F \mid i} (1-p_{F \mid i})} \right)^2}{ p_F - p_F^2 }. 
\end{align}
Notably, both ratios in Eqs. (\ref{Eq: Variance ratio 1} and \ref{Eq: Variance ratio 2}) do not depend on the sample size $N$, and the ratio $r_{ \frac{\text{SS,opt}}{\text{MCS}}}$ implies the optimal level of variance reduction of the stratified sampling compared to crude MCS. 
Eqs.~(\ref{Eq: Variance ratio 1}) and (\ref{Eq: Variance ratio 2}) reveal that the variance reduction achieved through stratification primarily stems from removing the variability within each stratum. 
Maximal variance reduction is achieved if $p_{F\mid I}$ is either 0 or 1 for each $i$, in which case, it holds that $r_{ \frac{\text{SS,prop}}{\text{MCS}}}=r_{ \frac{\text{SS,opt}}{\text{MCS}}}=0$.
By contrast, no variance reduction is achievable if $p_{F\mid I} = p_F$ holds for each $i$. 
These two conditions are not only sufficient but also necessary. 
For completeness, the proof of the latter statement is provided in Appendix A.

\begin{remark}
The analysis presented in this section does not rely on a particular stratification scheme. 
When the strata are selected in another way, $I$ should be viewed as an allocation variable, and $\lambda_i$ and $p_{F\mid i}$ denote the probability of the $i$-th stratum and the failure probability conditional on the $i$-th stratum, respectively.  
All the aforementioned conclusions remain valid.  
\end{remark}

\subsection{Implementation details}
\label{Subsec: implementation details of the classic stratified sampler}
To implement the stratified sampler in Eq.~(\ref{Eq: Classic stratified sampling}), one needs to compute the probability of the stratum $\lambda_i$ and an algorithm to sample from the conditional distribution $p_{\bm{X}(\bm{x}) \mid i}$, where $i=0,\cdots, n$. 
In the following, we present implementation details of the stratified sampling for two scenarios: one with independent and identically distributed (IID) components and the other with independent yet non-identically distributed (INID) components. 
In the first scenario, we reveal the inherent connection between the stratified sampling and the system signature, which, to the best of the authors' knowledge, has not been discussed previously; 
In the second scenario, we introduce the conditional Bernoulli model and its application in stratified sampling. 
Note that the analysis performed in Subsection \ref{Subsec: analysis of the classic stratified sampler} assumes that the sample size can be fractional. 
In practice, however, the sample size per stratum must be an integer and be at least one to maintain the unbiasedness of the stratified sampler. 
To address this, we propose a randomization strategy for the sample size. 
Additionally, the optimal sample allocation strategy requires knowledge of the conditional failure probability for each stratum. 
Since this information is unknown before the simulation, an approximation is necessary. 
In this work, the approximation is based on failure states and the assumption that the network performance is coherent. 

\subsubsection{Independent and identical components}
\label{Subsubsec: Independent and identical components}
The damage state $\bm{X}$ follows the IID multivariate Bernoulli distribution, whose probability mass function reads: 
\begin{equation} 
\label{Eq: IID Bernoulli distribution}
p_{\bm{X}}(\bm{x}) = p^{\sum_{i=0}^n x_i} (1-p)^{n-\sum_{i=0}^n x_i},
\end{equation}
where $p$ denotes the component failure probability, and $X_i=1$ indicates failure of the $i$-th component. 
% The state of each component is defined by a component-level performance function $g_{\text{comp}}(y)$, such that $X_i = \mathbb{I}\{g_{\text{comp}}(Y) \leq \gamma_i \}$. 
The number of failed components $I = \sum_{i=0}^n X_i$ therefore follows the binomial distribution, and $\lambda_i$ can be calculated as 
\begin{equation}
    \lambda_i = {n \choose i} p^i(1-p)^{n-i}. 
\end{equation} 
Given the number of failed components $i$, Eq.~(\ref{Eq: IID Bernoulli distribution}) is invariant with regard to which specific components failed. 
In other word, the conditional distribution $p_{\bm{X}\mid i}(\bm{x})$ is a uniform distribution residing in the set $\{\bm{x} \mid \sum_{i=1}^n x_i = i\}$. 
To sample from this distribution, one can randomly pick $i$ failed components without replacement and set the remaining components as safe. 

In this context, $p_{F\mid i}$ is closely related to the system signature \cite{Coolen&Coolen-Maturi2024}. 
The system signature, denoted as $\Phi_l$, is defined as the proportion of the system states with exactly $l$ functional components that also ensure the system functionality, relative to all such states regardless of system functionality. 
It then holds that: 
\begin{equation}
    \label{Eq: p_Fi for iid inputs}
    p_{F\mid i} = \frac{\mid \{ \bm{x} \mid \sum_{i=1}^n x_i = i, \bm{x} \in F \} \mid }{ {n \choose i} } = 1-\Phi_{n-i}, 
\end{equation}. 
The main motivation for deriving the system signature is the decoupling of the system evaluation from the component reliability. 
The latter changes over time but the system signature does not. 
Hence, the system reliability in function of time can be evaluated without recomputing the signature. 
Therefore, many sampling-based signature methods can be viewed as stratified sampling techniques with different sample allocation strategies. 
% For instance, Di Maio et al. \cite{DiMaio&others2023} proposed an entropy-driven Monte Carlo approach for approximating the survival signatures. 

\subsubsection{Independent but non-identical components} 
\label{Subsubsec: Independent but non-identical components}
For non-identical yet still independent components, the input distribution becomes 
\begin{equation}
p_{\bm{X}}(\bm{x}) = \prod_{i=1}^n p_i^{x_i} (1-p_i)^{1-x_i},
\end{equation}
where $p_i$ is the failure probability of the $i$-th component. 
In this context, the number of failed components $I = \sum_{i=1}^n X_i$ follows the Poisson-Binomial distribution, and $p_{\bm{X} \mid i}(\bm{x})$ is known as the conditional Bernoulli model \cite{Chen&Liu1997}. 
Therefore, algorithms for computing the probability mass function (PMF) of a Poisson-Binomial distribution and for generating samples from the conditional Bernoulli model \cite{Chen&others1994, Chen&Liu1997} can be directly applied to our purpose, that is, to calculate $\lambda_i=\Pr(I=i)$ and to sample from $p_{\bm{X} \mid i}(\bm{x})$. 

At the center of these methods lies the R function \cite{Chen&others1994}. 
Specifically, let $\mathcal{A}_{s}$ be a non-empty subset of $\mathcal{A} \triangleq \{1, \cdots, n\}$ and $1 \leq i \leq |\mathcal{A}_{s}|$, where $|\cdot|$ denote the cardinality of the set $|\mathcal{A}_{s}|$. 
The R function of $i$ and $\mathcal{A}_{s}$, denoted as $R(i, \mathcal{A}_{s})$, is defined as  
\begin{equation}
\label{Eq: R function}
R(i, \mathcal{A}_{s}) \triangleq \sum_{ \mathcal{B} \subset \mathcal{A}_{s}, |\mathcal{B}| = i} \left( \prod_{j \in \mathcal{B}} \frac{p_j}{1-p_j} \right), 
\end{equation}
with the convention $R(0, \mathcal{A}_{s}) = 1$ and $R(i, \mathcal{A}_{s}) = 0$ for any $i >|\mathcal{A}_{s}|$. 
The R function can be calculated efficiently through the recursive algorithm proposed in \cite{Gail&others1981, Chen&others1994}. 
We find that the probability $\lambda_i$ can subsequently be rewritten as
\begin{align}
    \lambda_i
    & = \sum_{\mathcal{B} \subset \mathcal{A}, |\mathcal{B}| = i} \left( \prod_{j \in \mathcal{B}} p_j  \prod_{j \notin \mathcal{B}} (1-p_j) \right) \notag \\
    & = \prod_{j \in \mathcal{A}} (1-p_j) \sum_{\mathcal{B} \subset \mathcal{A}, |\mathcal{B}| = i} \left( \prod_{j \in \mathcal{B}} \frac{p_j}{1-p_j} \right) \notag \\ 
    \label{Eq: stratum probability and R function}
    & \propto R(i, \mathcal{A}). 
\end{align}
The second equation follows the observation that, for any $\mathcal{B} \subset \mathcal{A}$, it holds that $\prod_{j \in \mathcal{A}} (1-p_j) = \prod_{j \in \mathcal{B}} (1-p_j) \prod_{j \notin \mathcal{B}} (1-p_j) $. 
Eq.~(\ref{Eq: stratum probability and R function}) shows that $\lambda_i$ can be determined through normalizing the R functions $R(i, \mathcal{A})$. 
Sampling from the conditional Bernoulli model $p_{\bm{X} \mid i}(\bm{x})$ can be accomplished through rejection sampling, but more efficient algorithms e.g., the ID-checking sampler \cite{Chen&Liu1997}, are advisable. 
The ID-checking sampler is summarized in Alg. \ref{Alg: ID-checking sampler}. 
Note that in line 4, we use Eq.(9) from \cite{Chen&Liu1997} to simplify the expression. 
\begin{algorithm}[!htbp]
\caption{The ID-checking sampler \cite{Chen&Liu1997}} 
\label{Alg: ID-checking sampler}
\LinesNumbered
\DontPrintSemicolon
\KwIn{The total number of components $n$, the number of failed components $i$, and the input distribution $p_{\bm{X}}(\bm{x})$ }
$\mathcal{B} \gets \emptyset$, $\bm{x}=(x_1=0,\cdots,x_n=0)$ \\
    \For{$k=1,\cdots,n$}{
        $r \gets |\mathcal{B}|$ \tcp*{$|\cdot|$ denotes the cardinality of the set $\mathcal{B}$}
        $\pi \gets \frac{R\left(i-r, \{k+1,k+2,\cdots,n\}\right)}{R\left(i-r, \{k,k+1,\cdots,n\}\right)}$ \tcp*{Function $R(\cdot,\cdot)$ is defined by Eq.(\ref{Eq: R function})}
        $u \sim \text{Uni}(0, 1)$ \tcp*{Sample $u$ uniformly from $(0, 1)$} 
        \uIf{$u > \pi$}{
            $\mathcal{B} \gets \mathcal{B} \cup \{k\}$\\
            $x_k \gets 1$ \\
        }
	}     
\KwOut{A sample $\bm{x} = (x_1,\cdots,x_n)$ following $p_{\bm{X}\mid i}(\bm{x})$}
\end{algorithm}
% \subsubsection{Common cause failures}

\subsubsection{Approximating the optimal sample allocation} 
\label{Subsubsec: Approximating the optimal sample allocation}
While implementing the proportional sample allocation to the strata is straightforward, the optimal sample allocation requires knowledge of the conditional failure probability within each stratum, $p_{F \mid i}$, which is unknown before simulation. 
A common strategy to address this issue is to launch a pilot run and estimate each $p_{F \mid i}$ through MCS within each stratum. 
However, when $p_{F \mid i}$ is small, MCS requires a large number of samples, making the pilot run computationally expensive and further reducing the efficiency of the final stratified sampler. 
Therefore, we propose a heuristic method to approximate these probabilities. 
% Therefore, we train a Bayesian additive regression tree (BART) as the surrogate model and substitute its prediction for the network performance evaluation during the pilot run. 

For connectivity-based metrics, we begin with a set of minimal cuts. 
Each minimal cut is an irreducible set of components whose failure can cause system failure. 
Here, 'irreducible' means that any subset of a minimal cut will not cause the system to fail. 
While the number of minimal cuts increases exponentially with the problem's dimension, identifying a subset thereof can be achieved through  stopping cut-finding algorithms such as \cite{Jasmon&Kai1985} prematurely. 
Since connectivity is a coherent metric, states that include any of the minimal cuts will also lead to system failure. 
The probabilities of these states can therefore be accumulated per stratum to obtain an approximation of the conditional failure probability. 
Specifically, let $\mathcal{C}_j$ denote the set of states that include the $j$-th minimal cuts, where $j$ ranges from 1 to $n_{\mathcal{C}}$, which denotes the number of available minimal cuts. 
Let $\mathcal{S}_i$ be the set of states that form the $i$-th stratum. 
An approximation $\widetilde{p}_{F \mid i}$ of the conditional failure probability of the $i$-th stratum can be obtained as
\begin{equation}
\widetilde{p}_{F \mid i} = \sum_{ \bm{x} \in \mathcal{X}_i} p_{\bm{X}}(\bm{x}), \quad \mathcal{X}_i = \bigcup_{j=1}^{n_\mathcal{C}} \left(\mathcal{S}_i \cap \mathcal{C}_j\right). 
\end{equation}
If the states in $\mathcal{X}_i$ are too numerous to count, one could alternatively set the minimum of one and $\sum_{i=1}^{n_\mathcal{C}} \Pr(\mathcal{S}_i \cap \mathcal{C}_j)$. 
Note that $\widetilde{p}_{F \mid i}$ is only used in Eq.~(\ref{Eq: The optimal budget allocation strategy}) for approximating the optimal sample size, that is:  
\begin{equation}
\label{Eq: The near-optimal budget allocation strategy}
    N_i^{(\text{aopt})} \triangleq N \frac{ \lambda_i \sqrt{\widetilde{p}_{F \mid i} (1-\widetilde{p}_{F \mid i})} }{ \sum_{k=0}^{n}\lambda_k \sqrt{\widetilde{p}_{F \mid k} (1-\widetilde{p}_{F \mid k})} }, \quad i=0,\cdots,n,
\end{equation}
where $N_i^{(\text{aopt})}$ denotes the approximation of the optimal sample size. 
% The approximation error can be small if $\widetilde{p}_{F \mid i}$ consistently underestimates the true value, $p_{F \mid i}$, across strata.  

Physics-based system performance functions, however, are not always coherent, and efficient minimal-cut-finding algorithms may not be available. 
Nevertheless, we proceed as if they were coherent. 
We begin with a set of failure states where no state is larger than any other. 
Note that, unlike a minimal cut, which is a vector that includes only failed components, the failure state collects the states of each component, with 1 denoting failure and 0 functional state. 
Besides, state $\bm{x}$ is larger than $\bm{y}$ if $x_i \geq y_i$ for each $i$-th component in $\bm{x}$. 
These failure states can be obtained as a by-product of genetic algorithm (GA), which is initially designed to remove redundant strata in Section \ref{Subsec: Overall workflow}. 
States larger than any of the observed failure states are also assumed to cause failure, and their probabilities are accumulated per stratum for estimating the conditional failure probability, in a similar manner as for connectivity-based metrics. 
These probabilities are then inserted into Eq.~(\ref{Eq: The optimal budget allocation strategy}) for approximating the optimal sample size within each stratum, which is subsequently randomized to a neighboring integer according to Eq.~(\ref{Eq: randomized sample size}). 

While the proposed heuristic approximation introduces errors in approximating $p_{F \mid i}$, the optimal sample size is not significantly affected by these errors. 
The relative increase in variance resulting from the approximation, denoted as $\alpha$, is a weighted average of the squared relative differences in the sample size following Eq.~(\ref{Eq: the relative variance increase due to a non-optimal sample allocation strategy}). 
In particular, let $\{N_i\}_{i=0}^n$ denote a sample allocation, e.g., $\{N^{(\text{aopt})}_i\}_{i=0}^n$, and $\{N_i^{(\text{opt})}\}_{i=0}^n$ be the optimal sample allocation with the same sample size $N$, i.e., $\sum_{i=1}^n N_i = \sum_{i=1}^n N_i^{(\text{opt})} = N$. 
The variances of the subsequent stratified samplers with $\{N_i\}_{i=0}^n$ and $\{N_i^{(\text{opt})}\}_{i=0}^n $ are denoted as $\mathbb{V}\left(\widehat{p}_F^{(\text{SS})}\right)$ and $\mathbb{V}\left(\widehat{p}_F^{(\text{SS,opt})}\right)$, respectively. 
It can be proven that \cite{Cochran1977}:
\begin{equation}
    \label{Eq: the relative variance increase due to a non-optimal sample allocation strategy}
    \alpha \triangleq \frac{ \mathbb{V}\left(\widehat{p}_F^{(\text{SS})}\right) - \mathbb{V}\left(\widehat{p}_F^{(\text{SS,opt})}\right) }{\mathbb{V}\left(\widehat{p}_F^{(\text{SS,opt})}\right)} = \sum_{i=0}^n \frac{N_i}{ \sum_{i=0}^n N_i} \left( \frac{ N_i - N_i^{(\text{opt})}}{ N_i } \right)^2. 
\end{equation}
% Eq.~(\ref{Eq: the relative variance increase due to a non-optimal sample allocation strategy}) states that the relative increase in variance is not sensitive to the relative difference in the sample size given the relative difference is smaller than one. 
This implies that approximating $N_i^{(\text{opt})}$ will not lead to a significant variance increase as long as there is no $N_i$ that is significantly smaller than $N_i^{(\text{opt})}$. 
In fact, it is evident that: 
\begin{equation}
    \alpha \leq \left( \max_{i=0,\cdots,n} \frac{ \vert N_i - N_i^{(\text{opt})} \vert }{ N_i } \right)^2. 
\end{equation}

\subsubsection{Randomizing the sample size}
\label{Subsubsec: Randomizing the sample size} 
In practice, the sample size per stratum should be an integer and at least one to maintain the unbiasedness of the stratified sampler. 
However, given an initial sample size of $N$, if the allocation strategy does not inherently guarantee an integer sample size for each $i$-th stratum, $N_i$, we need to randomize the sample size to be either the floor integer $\lfloor N_i \rfloor$ or the ceiling integer $\lceil N_i \rceil$. 
Specifically, let $\overline{N_i}$ denote the randomized sample size of the $i$-th stratum. 
It follows the following Bernoulli distribution: 
\begin{equation}
\label{Eq: randomized sample size}
    \overline{N_i} = 
    \begin{cases}
        \lfloor N_i \rfloor & \text{with prob.} \quad \frac{\lfloor N_i \rfloor \lceil N_i \rceil}{N_i} - \lfloor N_i \rfloor \\
        \lceil N_i \rceil & \text{with prob.} \quad \lceil N_i \rceil - \frac{\lfloor N_i \rfloor \lceil N_i \rceil}{N_i}.
    \end{cases}
\end{equation}
It is evident that $ \mathbb{E} \left[ \frac{1}{\overline{N_i}} \right] = \frac{1}{N_i}$ when $N_i>1$, and $\overline{N_i}=1$ when $0 < N_i < 1$. 
When $N_i>1$ for each $i$, this ensures an unbiased stratified estimator with expected variance consistent with Eq.~(\ref{Eq: The variance of the classic stratified sampling estimator}). 
Consequently, Eqs.~(\ref{Eq: Variance ratio 1}) and (\ref{Eq: Variance ratio 2}) can be interpreted as the expected variance ratios averaged over all randomized sample sizes. 
On the other hand, if many strata have a sample size $N_i < 1$, the actual computational cost, which equals $\sum_i \overline{N_i}$, can significantly exceed the initial sample size $N$, because sampling from Eq.~(\ref{Eq: randomized sample size}) will result in a single sample for all such strata. 

In summary, to implement optimal sample allocation with an initial sample size $N$, we estimate $\widetilde{p}_{F \mid i}$ for each $i$-th stratum as described in Subsection \ref{Subsubsec: Approximating the optimal sample allocation}, approximate the optimal sample size $N_i^{(\text{aopt})}$ with Eq.~(\ref{Eq: The optimal budget allocation strategy}), and then randomize each $N_i^{(\text{aopt})}$ into a neighbouring integer $\overline{N_i^{(\text{aopt})}}$ using Eq.~(\ref{Eq: randomized sample size}). 
The resulting stratified sampler is denoted as $\widehat{p}_F^{(\text{SS},\overline{\text{aopt}})}$. 
Next, we propose two improvements to $\widehat{p}_F^{(\text{SS},\overline{\text{aopt}})}$: one involves eliminating redundant strata (Section 3), and the other introduces a refinement procedure (Section 4). 
% Alternatively, one may consider utilizing the mixture distribution $\sum_{i=0}^n \omega_i p_{\bm{X} \mid i}(\bm{x})$ for conducting importance sampling, where $\omega_i$ is the weight of the $i$-th mixture component, subject to the constraint $\sum_{i=0}^n \omega_i = 1$. 
% This estimator is unbiased with a variance $\sum_{i=0}^n \frac{\lambda^2_i}{\omega_i} p_{F \mid i} - p^2_F$. 
% It is evident that by setting $\omega_i = \lambda_i $, the variance of this importance sampling estimator is the same as that of crude MCS, and the optimal selection of $\omega_i$ to minimize the variance is proportional to $\lambda_i \sqrt{p_{F \mid i}}$.

\section{Removing redundant strata} 
\label{Sec: Removing redundant strata}
\subsection{Stratified sampler with redundant strata removed}
\label{Subsec: Stratified sampler after excluding the redundant strata}
The network is always in safe state when no components fail. 
Similarly, if one can identify the minimum number of failed components required to cause network failure, denoted as $i^*$, any stratum with fewer than $i^*$ failed components can be removed, since the conditional probability of failure of such a stratum will be zero. 
% Samples are generated only within the remaining strata. 
Eq.~(\ref{Eq: Total probability theorem}) then becomes:  
\begin{subequations}
    \begin{equation}
        p_F = p_F^* \sum_{i=i^*}^n \lambda_i, 
    \end{equation}
    \begin{equation}
    \label{Eq: constraint failure probability}
        p_F^* = \sum_{i=i^*}^n  \frac{\lambda_i}{\sum_{j=i^*}^n \lambda_j} p_{F \mid i}.
    \end{equation}
\end{subequations}
Here, $p_F^*$ is the failure probability conditional on $I \geq i^*$, i.e., $\Pr(F \mid I\geq i^*)$, and can be significantly larger than $p_F$. 
Hence, one can estimate $p^*_F$, using either crude MCS or stratified sampling, and then scale the result to $p_F$ by multiplying it with $\sum_{j=i^*}^n \lambda_j$, which can be calculated exactly. 
In other words, we investigate the following two estimators: 
\begin{align}
    \label{Eq: conditional MCS estimator}
    & \widehat{p}_F^{(\text{cMCS})} =  \left( \sum_{j=i^*}^n \lambda_j \right) \frac{1}{N} \sum_{k=1}^N \mathbb{I} \{ \bm{x}_k \in F\},  \quad \bm{x}_k \sim p_{\bm{X} \mid I \geq i^*}(\bm{x}) \\
    \label{Eq: conditional stratified sampler}
    \widehat{p}_F^{(\text{cSS})} &= \left( \sum_{j=i^*}^n \lambda_j \right) \sum_{i=i^*}^n \frac{\lambda_i}{\sum_{j=i^*}^n \lambda_j} \frac{1}{N_i} \sum_{k=1}^{N_i} \mathbb{I}\{ \bm{x}_k \in F\}, \quad \bm{x}_k \sim p_{\bm{X} \mid i}(\bm{x}).
\end{align}
We refer to the first estimator, $\widehat{p}_F^{(\text{cMCS})}$, as conditional MCS, since it generates samples that possess at least $i^*$ failed components, following the truncated input distribution, $p_{\bm{X} \mid I\geq i^*}(\bm{x}) \propto p_{\bm{X}}(\bm{x}) \mathbb{I}\{ \sum_{d=1}^n x_d \geq i^*\}$. 
This can be achieved through methods such as rejection sampling. 
By contrast, crude MCS samples in the unconstrained input space, following input distribution $p_{\bm{X}}(\bm{x})$. 
The second estimator, denoted as $\widehat{p}_F^{(\text{cSS})}$, is similar to Eq.~(\ref{Eq: Classic stratified sampling}) but excludes the strata where the states have fewer than $i^*$ failed components. 
It is termed the conditional stratified sampler. 

\subsection{Identifying $i^*$ for connectivity-based performance function}
\label{Subsec: determine the i_star for connectivity-based performance metrics}
We observe that $i^*$ represents the cardinality of the minimum cut, that is, a minimal cut with the smallest number of failed components, and can be efficiently determined for specific performance metrics. 
As an example, the cardinality of the minimum cut set that disconnects the source $\text{s}$ and the sink $\text{t}$ equals the maximum flow from $\text{s}$ to $\text{t}$, assuming unit capacity for each edge\footnote{Although the Ford-Fulkerson algorithm is originally proposed for directed networks, the adaptation to undirected networks is straightforward.} \cite{Ford&Fulkerson1956}. 
If there are more than two terminal nodes, $i^*$ can be identified by examining the minimal-cardinality $\text{s}-\text{t}$ cuts for each pair of terminal nodes. 
The number of required maximum flow analyses is therefore equal to ${K}\choose{2}$, where $K$ denotes the number of terminal nodes.  
For all-terminal connectivity, more efficient algorithms such as the Soter and Wagner algorithm \cite{Stoer&Wagner1997} can be employed for identifying $i^*$. 

\subsection{Identifying $i^*$ for physics-based performance metrics} 
\label{Subsec: determine the i_star for physics-based performance metrics}
For physics-based metrics, determining $i^*$ generally requires solving the following optimization problem: 
\begin{align}
\label{OPT: the critical number of failed components}
i^* = \min\limits_{i=0,1,\cdots,n} & \; i\\
\text{s.t.} & \; p_{F \mid i} \neq 0 \notag
\end{align}
The optimization problem can be reformulated as follows: 
\begin{align}
\label{OPT: the critical number of failed components (reformulated)}
i^* = \min\limits_{\bm{x} \in \{0,1\}^n } & \; \sum_{d=1}^n x_d\\
\text{s.t.} & \; \bm{x} \in F \notag
\end{align}
where $\bm{x} = (x_1, \cdots, x_n) \in \{0, 1\}^n$ represents the binary system state, and $\gamma$ is the threshold for defining network failure. 
Note that the performance function $g(\bm{x})$ for defining the failure event $F$ is not necessarily convex after continuous relaxation of discrete variables $\bm{x}$, i.e., allowing $\bm{x}$ to take values from $[0,1]^n$ instead of $\{0,1\}^n$. 
% For example, the maximum flow through a series system with edge capacity specified by $\bm{x}$ depends on the minimum capacity of all edges,i.e., $g(\bm{x}) = \min_d (x_d)$, which is not convex. 
In some cases, the performance function $g(\bm{x})$ is a black-box function. 
Consequently, most relaxation- and factorization-based methods are not applicable \cite{Burer&Letchford2012}, and heuristic evolutionary methods should be employed. 
% Broadly speaking, optimization (\ref{OPT: the critical number of failed components (reformulated)}) belongs to the mixed integer nonlinear programming (MINP) problem. 
% A survey of methodologies for solving MINP is given in (Burer \& Letchford 2012). 
% Major concepts: 
% (1) continuous relaxation, e.g., $x \in {0,1}$ is relaxed as $ x \in \rightarrow [0,1]$. 
% (2) convex(linear) relaxation, which depends on finding appropriate convex(linear) under-estimating functions and/or concave(linear) over-estimating functions.
% (3) branch-and-bound: the basic idea is to recursively decompose the original optimization problem into a sequence of subproblems that are easier to solve. 
% It is initially employed for mixed-integer linear programming when the optimal solution after continuous relaxation is not optimal. 
% For mixed-integer non-linear programming problems, further relaxation techniques such as convex relaxation are often required. 
% (4) factorization facilitates the constraint function's convex or linear relaxation by substituting additional variables and constraints. 

Genetic algorithm (GA) is one of the most popular evolutionary algorithms, drawing inspiration from natural selection \cite{Mitchell1998}. 
The algorithm exhibits broad applicability, often performs well in low to moderate dimensions, and is simple to implement, making it well-suited for solving (\ref{OPT: the critical number of failed components (reformulated)}) with intricate performance functions. 
The main process involves initializing a set of encoded solutions called chromosomes and evolving them toward better solutions through selection, crossover (or recombination), mutation, and other nature-inspired operators. 
% These operators are essential to the efficiency of GA. 
As (\ref{OPT: the critical number of failed components (reformulated)}) only involves binary solutions, it is natural to encode them into binary strings of bits, either 0 or 1. 
% Also, we did not observe significant outperformance of the real-encoded genetic algorithm in our numerical experiments. 

We start with $n_{\text{pop}}$ chromosomes, each generated from an independent Bernoulli distribution. 
The population size $n_{\text{pop}}$ plays an essential role in the efficiency of GA. 
A large value facilitates finding the global minimum, albeit with a larger computational cost. 
To filter out the 'elite' chromosomes with lower objective values, $n_{\text{trn}}$ candidates are randomly selected with replacement from the previous population, and the one with the smallest objective function value is kept. 
This is known as the tournament selection. 
In contrast to other selection techniques, such as Routte Wheel selection, tournament selection does not require scaling the objective function but depends on the tournament size $n_{\text{trn}}$. 
A large value of $n_{\text{trn}}$ leads to rapid yet frequently premature convergence of the algorithm. 
% The nonlinear constraint is incorporated into the objective function through a penalty term. 
After getting the 'elite' chromosomes from tournament selection, we enrich their diversity through the uniform crossover and uniform mutation operators. 
Both operations are straightforward for binary-encoded chromosomes. 
In the uniform crossover, each bit of the two chromosomes is exchanged with a probability of 0.5, and in the uniform mutation, each bit has a probability $p_{\text{mt}}$ to flip. 
% Note that an average of $n\cdot p_{\text{mt}}$ bits are flipped during the mutation, where $n$ is the dimension. 
% Consequently, a large $p_{\text{mt}}$ often introduces less promising chromosomes with many failed components. 
The crossover fraction, denoted as $f_{\text{xo}}$, is another important parameter that governs the proportion of chromosomes undergoing the crossover and mutation. 
For instance, selecting $f_{\text{xo}} = 0.8$ indicates that, in each generation, the number of chromosomes undergoing crossover is four times greater than those undergoing mutation. 
A parameter study of GA is undertaken in the numerical examples in Section \ref{Sec: Numerical Examples}. 
% Investigations over more advanced and tailed operators and other potential evolutionary algorithms are beyond the scope of this paper and will be left to future research. 

% A comprehensive investigation of GA and other potential evolutionary algorithms is beyond the scope of this paper and will be left to future research. 
When employing these heuristic algorithms, there is no assurance of locating the global minimum $i^*$ with a limited computational budget, and the algorithms may terminate prematurely, resulting in a local minimum larger than $i^*$. 
Consequently, a bias will be introduced to the subsequent stratified sampler. 
Alternative to solving the optimization problem of Eq.~(\ref{OPT: the critical number of failed components (reformulated)}), one can enumerate the states with a small $i$, e.g., for $i \leq 3$. 
In the worst case, a lower bound of $i^*$, denoted as $\underline{i^*}$, can be obtained. 
Sampling conditional on $\sum_{d=1}^n x_d \geq \underline{i^*}$ ensures an unbiased stratified sampler, yet its efficiency declines when $\underline{i^*}$ is significantly below $i^*$. 

\section{Stratum refinement}
\label{Sec: stratum refinement}
\subsection{Stratified sampler with unbalanced refinement}
The performance of the conditional stratified sampler $\widehat{p}_F^{(\text{cSS})}$ can be further enhanced through iteratively refining the remaining $n-i^*+1$ non-redundant strata that consist of states having at least $i^*$ failed components. 
Suppose after $T$ iterations of the refinement procedure, each $i$-th stratum is split into $n_i$ sub-strata, $\mathcal{S}_{i, j}$, with their probabilities (or sizes) denoted as $\lambda_{i,j}$, where $i = i^*,\cdots,n$ and $j=1,\cdots,n_i$. 
In total $n_T = \sum_{i=i^*}^n n_i$ refined strata are generated. 
We propose a refinement stratified sampler, denoted as $\widehat{p}_F^{(\text{SSuR})}$, that reads as follows: 
\begin{align}
\label{Eq: refined stratified sampler}
    \widehat{p}_F^{(\text{SSuR})} 
    & = \left( \sum_{l=i^*}^n \lambda_l \right) \sum_{i=1}^n \sum_{j=1}^{n_i} \frac{\lambda_{i,j}}{\sum_{l=i^*}^n \lambda_l} \frac{1}{N_{i,j}} \sum_{k=1}^{N_{i,j}} \mathbb{I}\{ \bm{x}_k^{(i,j)} \in F\}, \\
    & \bm{x}_k^{(i,j)} \sim p_{i,j}(\bm{x}) \propto p_{\bm{X}}(\bm{x}) \mathbb{I}\{\bm{x} \in \mathcal{S}_{i,j} \}, \notag
\end{align}
where $N_{i,j}$ denotes the sample size of the stratum $\mathcal{S}_{i,j}$, and $p_{i,j}(\bm{x})$ is the sampling distribution. 
In particular, each sub-stratum $\mathcal{S}_{i,j}$ is uniquely characterized by a partition of network components and the number of failed components within the different clusters of that partition. 
For independent components, the conditional Bernoulli model is borrowed to calculate $\lambda_i$ and to sample from $p_{i,j}(\bm{x})$. 
For determining $N_{i,j}$, we consider three sample allocation strategies: proportional, optimal, and practical (approximation+randomization) allocation. 
% The first two strategies are introduced for analtyical analysis, and the last one is designed for practical implementation. 
The resulting estimators are denoted as $\widehat{p}_F^{(\text{SSuR,prop})}$, $\widehat{p}_F^{(\text{SSuR,opt})}$, and $\widehat{p}_F^{(\text{SSuR},\overline{\text{aopt}})}$, respectively. 

As the stratum is not necessarily split equally at each iteration, we term our method \emph{stratified sampling with unbalanced refinement} (SSuR), also to distinguish it from the refined stratified sampler proposed in Shields et al. \cite{Shields&others2015}. 
In \cite{Shields&others2015}, a stratum is divided into multiple equal-sized sub-strata, with each sub-stratum assigned a single sample in the final estimator. 
This is known as the balanced refinement. 
In contrast, our approach is specifically designed for network reliability problems, where the inputs are often discrete, making it challenging or impossible to divide a stratum equally. 
We therefore introduce an unbalanced refinement procedure. 
In addition, we focus on optimal sample allocation instead of the one-sample-per-stratum strategy. 

The computational cost of $\widehat{p}_F^{(\text{SSuR})}$ is measured by the number of network performance evaluations, including those for determining $i^*$ and for stratified sampling. 
Under the same computational cost, the variance ratios of $\widehat{p}_F^{(\text{SSuR})}$ over conditional MCS can be expressed as follows: 
\begin{align}
    \label{Eq: Variance ratio 1, SS vs cMCS}
    r_{ \frac{\text{SSuR,prop}}{\text{cMCS}}} 
    & = \frac{\mathbb{V}\left( \widehat{p}_F^{(\text{SSuR,prop})} \right)}{\mathbb{V}\left( \widehat{p}_F^{(\text{cMCS})} \right)}  = \frac{ \sum_{i=i^*}^n \sum_{j=1}^{n_i} \frac{\lambda_{i,j} }{\sum_{l=i^*}^n \lambda_l} p_{F \mid i,j} (1-p_{F \mid i,j}) }{ p^*_F - (p^*_F)^2},\\ 
    \label{Eq: Variance ratio 2, SS vs cMCS}
    r_{ \frac{\text{SSuR,opt}}{\text{cMCS}}} & = \frac{\mathbb{V}\left( \widehat{p}_F^{(\text{SSuR,opt})} \right)}{\mathbb{V}\left( \widehat{p}_F^{(\text{cMCS})} \right)} = \frac{\left( \sum_{i=i^*}^{n} \sum_{j=1}^{n_i} \frac{\lambda_{i,j}}{\sum_{l=i^*}^n \lambda_l} \sqrt{p_{F \mid i,j} (1-p_{F \mid i,j})} \right)^2}{ p^*_F - (p^*_F)^2 },
\end{align}
where $p_{F \mid i,j}$ denotes the conditional failure probability of the stratum $\mathcal{S}_{i,j}$. 
Note that Eqs.~(\ref{Eq: Variance ratio 1, SS vs cMCS}) and (\ref{Eq: Variance ratio 2, SS vs cMCS}) take the same form as Eqs.~(\ref{Eq: Variance ratio 1}) and (\ref{Eq: Variance ratio 2}), with $\lambda$ and $p_F$ replaced by the normalized size $\frac{\lambda_{i,j}}{\sum_{i=i^*}^n \lambda}$ and $p^*_F$, respectively, and excluding the redundant strata. 
% The two ratios $r_{ \frac{\text{SSuR,prop}}{\text{cMCS}}}$ and $r_{ \frac{\text{SSuR,opt}}{\text{cMCS}}}$ are independent of the initial sample size $N$ and of the cost of computing $i^*$, denoted as $N'$.  

\subsection{Stratum refinement never increases the variance ratio} 
\label{Subsec: monotonicity property of stratum refinement}
% The idea of the stratum refinement is motivated by the rapid decay of $\lambda_i$ as $i$ goes larger. 
% Consequently, when adopting the proportional allocation strategy, a large portion of samples are located in the first few strata. 
% In the extreme case, all samples fall in the $i^*$-th strata, and the stratified sampling becomes just crude MCS. 
% Therefore, it is natural to refine the strata with large probabilities, i.e., $\lambda_i$.
% To minimize the variance of the stratified sampler, the strata should be selected such that the conditional failure probability on each stratum approaches either zero or one. 
% However, this requires prior knowledge of the limited state function, which is often inaccessible before simulation. 
The rationale behind the stratum refinement stems from the observation that the variance ratios $r_{ \frac{\text{SSuR,prop}}{\text{cMCS}}}$ and $r_{ \frac{\text{SSuR,opt}}{\text{cMCS}}}$ (as well as $r_{ \frac{\text{SS,prop}}{\text{MCS}}}$ and $r_{ \frac{\text{SS,opt}}{\text{MCS}}}$) do not increase when further splitting any stratum, $\mathcal{S}_{i,j}$, into two sub-strata $\mathcal{S}_{i,j_1}$ and $\mathcal{S}_{i,j_2}$. 
This property is also observed by Pettersson and Krumscheid \cite{Pettersson&Krumscheid2022}.   
To further enhance clarity, we provide an independent and more detailed proof in Appendix B. 
Specifically, we prove that, when fractional sample sizes are permitted, it follows that: 
\begin{subequations}
\begin{equation}
    \label{Eq: rationale of stratum refinement 1}
     \lambda_{i,j_1} \cdot p_{F\mid i,j_1}(1-p_{F\mid i,j_1}) + \lambda_{i,j_2} \cdot p_{F\mid i,j_2}(1-p_{F\mid i,j_2}) - \lambda_{i,j} \cdot p_{F\mid i,j}(1-p_{F\mid i,j}) \leq 0, 
\end{equation} 
\begin{equation}
    \label{Eq: rationale of stratum refinement 2}
    \lambda_{i,j_1} \sqrt{p_{F\mid i,j_1}(1-p_{F\mid i,j_1})} + \lambda_{i,j_2} \sqrt{p_{F\mid i,j_2}(1-p_{F\mid i,j_2})} - \lambda_{i,j} \sqrt{p_{F\mid i,j}(1-p_{F\mid i,j})} \leq 0. 
\end{equation}
\end{subequations}
% where $\lambda_{i_1}$ and $\lambda_{i_2} $ denote the probabilities of the two sub-strata, and $p_{F\mid i_1}$ and $p_{F\mid i_2}$ denote the conditional failure probabilities within the sub-strata. 
By substituting Eqs.~(\ref{Eq: rationale of stratum refinement 1}) and (\ref{Eq: rationale of stratum refinement 2}) into Eqs.~(\ref{Eq: Variance ratio 1, SS vs cMCS}) and (\ref{Eq: Variance ratio 2, SS vs cMCS}), respectively, it is evident that refining strata will not increase $r_{ \frac{\text{SSuR,prop}}{\text{cMCS}}}$ and $r_{ \frac{\text{SSuR,opt}}{\text{cMCS}}}$. 
The proof is independent of how the strata are refined. 
When the discrete sample space is completely stratified with each sample state forming a stratum, the variance ratios $r_{ \frac{\text{SSuR,prop}}{\text{cMCS}}}$ and $r_{ \frac{\text{SSuR,opt}}{\text{cMCS}}}$ become zero. 
% In practice, $N_i,j$ cannot be fractional, we therefore approximate the optimal sample size and randomize it to an neighbouring integer as demonstrated in Subsection \ref{Subsec: implementation details of the classic stratified sampler}. 
% If $N_i \geq 1$ holds for each $i$-th stratum, the monotonicity property of the refinement remains preserved for the expectation of the variance ratios averaged across randomized sample sizes. 
% By contrast, if $N_i<1$ holds for any stratum, this statum will be assigned a single sample. 

\subsection{Refinement procedure} 
\label{Subsec: one-step refinement procudure}
\begin{figure}[htbp]
    \centering
    \includegraphics[scale=0.7]{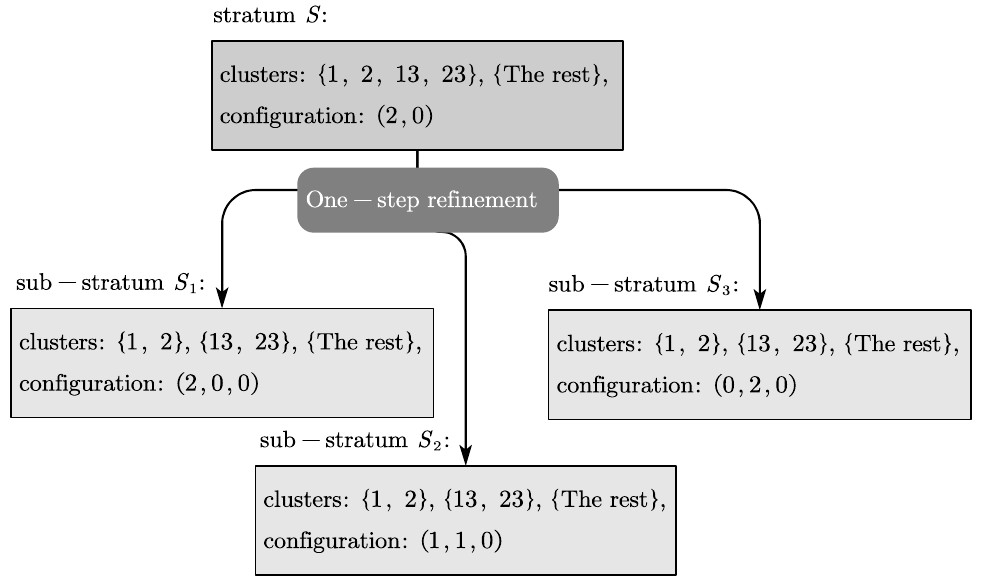}
    \caption{A schematic plot of the refinement procedure.}
    \label{Fig: A schematic plot of the refinement procedure}
\end{figure}

The proposed refinement procedure involves characterizing each stratum by partitioning its components and counting the number of failed components within each cluster of the partition. 
The components in a cluster are not necessarily of the same type. 
Different clusters are mutually exclusive and collectively exhaustive. 
For instance, the stratum $S$ in Fig.~\ref{Fig: A schematic plot of the refinement procedure} is defined by having two failed components in components 1,2, 13, and 23, and having no failed components in the rest. 
This stratum has two clusters: one containing components 1, 2, 13, and 23, and the other containing the remaining components. 
The number of failed components within each cluster equals 2 and 0, respectively. 
% We use the vector $(2, 0)$ to represent this configuration of failed components. 

For independent components, each cluster, along with the number of failed components within the cluster, follows a conditional Bernoulli distribution. 
Therefore, calculating its probability (size) or performing conditional sampling within each cluster is straightforward by using the R function described in Subsection \ref{Subsubsec: Independent but non-identical components}. 
The component states within different clusters are independent. 
Thus, the stratum size is the product of the sizes of all its clusters, and conditional sampling within the stratum can be done cluster by cluster. 
The conditional failure probability of each stratum can be viewed as a generation of the failure signature where the components in each cluster may vary in type. 

One principle of stratum refinement is to reduce the variability of the stratum size. 
To achieve this, we split the most probable cluster within the most probable stratum, i.e., the cluster and the stratum with the largest probability. 
Specifically, suppose we want to refine the stratum $S$ in Fig.~\ref{Fig: A schematic plot of the refinement procedure}. 
Note that there are no failed components in its second cluster, and splitting this cluster into smaller groups leads to no improvement. 
Therefore, we keep the second and split the first cluster, resulting in two sub-clusters: one containing components 1 and 2, and the other containing components 13 and 23. 
The number of failed components in the first cluster, which is two, is then distributed between the two sub-clusters, and there are three possible configurations: $(2, 0)$, $(1, 1)$, and $(0, 2)$. 
Here, the configuration $(2, 0)$ indicates that two components in the first cluster have failed, while all components in the second cluster remain functional. 
Consequently, three new sub-strata are generated during this decomposition, each having the same partition, or division of clusters, but with different configurations of failed components within each cluster. 
% In principle, refining a cluster with $j$ failed components leads to $j+1$ sub-strata. 
A schematic illustration of the refinement procedure is provided in Fig.~\ref{Fig: A schematic plot of the refinement procedure}. 

\subsection{Number of refinement steps, $T$}
In principle, one should refine the strata as much as possible since this will not increase but often decrease the variance ratio. 
Ultimately, when each state constitutes a stratum, the conditional probability will be either 0 or 1, resulting in a zero-variance estimator. 

In practice, however, the optimal sample size is unknown and it has to be an integer, so we apply the approximation and randomization proposed in Subsection \ref{Subsubsec: Randomizing the sample size}, which depends on an initial sample size $N$ and a division of the strata. 
The total computational cost of the proposed estimator $\widehat{p}_F^{(\text{SSuR},\overline{\text{aopt}})}$ consists of two parts: the cost for determining $i^*$, denoted as $N^{(\text{plt})}$, and the cost for stratified sampling, which equals $\sum_{i,j} \overline{N_{i,j}^{(\text{aopt})}}$. 
Here, $\overline{N_{i,j}^{(\text{aopt})}}$ is the sample size for the stratum $\mathcal{S}_{i,j}$ after approximation and randomization. 
The first part of the cost is independent of $N$ and $T$. 
However, the second part of the cost gradually increases with $T$ under a fixed $N$. 
This is because, as $T$ becomes larger, more strata will have zero conditional failure probabilities, and therefore will be assigned a single sample when using Eq.~(\ref{Eq: randomized sample size}). 
In fact, the stratified sampler gradually degenerates to brute-force enumeration over increasing refinement steps $T$. 

Given a fixed computational budget, one could set the initial sample size $N$ to a certain portion of the budget and then increase the number of refinement steps until the actual cost first exceeds the fixed budget. 
Such tuning of hyper-parameters requires no additional evaluation of the network performance function. 
A more sophisticated approach to determining the number of refinement steps involves using adaptive stratification \cite{Etore&others2011}, where an approximation of the estimator's variance is minimized through jointly sampling and adapting the strata and sample allocation. 
If this variance approximation is less than a specified threshold, the refinement steps are terminated. 
The robustness and effectiveness of this method in network reliability assessment are left to future work. 

\subsection{Overall workflow}
\label{Subsec: Overall workflow}
\begin{figure}[htbp]
    \centering
    \includegraphics[scale=0.8]{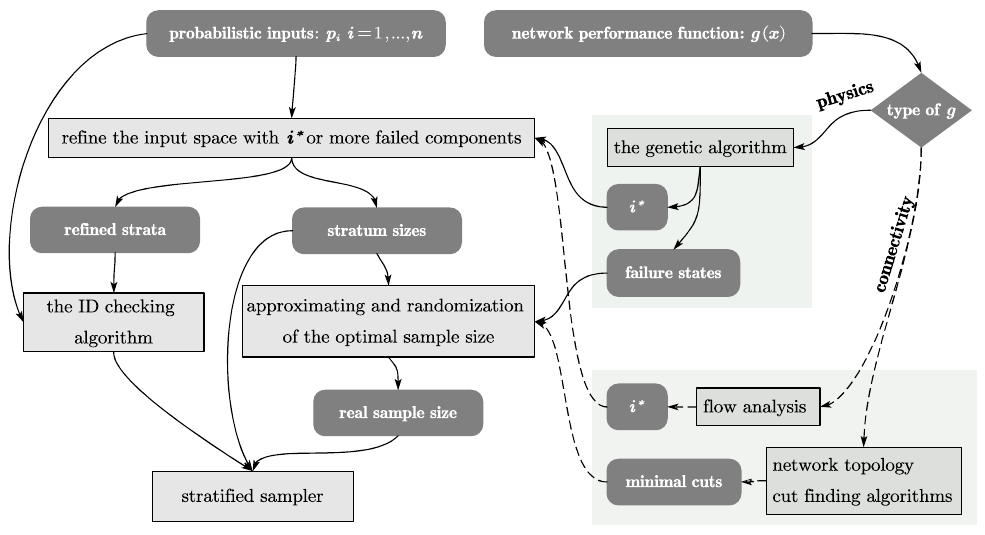}
    \caption{Workflow of the proposed stratified sampler.}
    \label{Fig: workflow of the proposed stratified sampler}
\end{figure}

The overall workflow is depicted in Fig.~\ref{Fig: workflow of the proposed stratified sampler}. 

One first proceeds by identifying the minimum cut's cardinality $i^*$ to eliminate the redundant strata containing less than $i^*$ failed components. 
Subsequently, one iteratively refines the remaining $n-i^*+1$ strata and approximates the optimal sample size for each refined stratum. 
These approximated sample sizes are then randomized to a neighbouring integer, as described in Eq.~(\ref{Eq: randomized sample size}), to facilitate the final stratified sampler. 

For connectivity-based performance metrics, $i^*$ can be computed as a by-product of the maximum flow analysis (See Subsection \ref{Subsec: determine the i_star for connectivity-based performance metrics}). 
After stratum refinement, the conditional failure probability of each stratum is approximated using a set of minimal cuts obtained through either network topology or tailored cut-finding algorithms \cite{Yeh2008}. 
Since connectivity is a coherent metric, states that include at least one minimal cut will also lead to system failure. 
The probability of these states is then accumulated per stratum as the estimated conditional failure probability of each stratum. 
For more details, we refer to Subsection \ref{Subsubsec: Approximating the optimal sample allocation}. 

For physics-based performance metrics, $i^*$ is estimated using GA (See Subsection \ref{Subsec: determine the i_star for physics-based performance metrics}). 
During this process, each generated individual and its corresponding network performance are recorded, and the individuals causing the system failure are identified. 
Assuming coherency of the network performance metric, the conditional failure probabilities can be estimated similarly to the connectivity metric, by using the failure-inducing individuals. 
For more details, we refer to Subsection \ref{Subsubsec: Approximating the optimal sample allocation}. 

\section{Numerical Examples}
\label{Sec: Numerical Examples}
We present two numerical examples to demonstrate the efficiency of the SSuR estimator with practical sample allocation i.e., $\widehat{p}_F^{(\text{SSuR},\overline{\text{aopt}})}$: one related to the power flow analysis and the other to the connectivity of a water supply system. 
The efficiency of an estimator is defined as being inversely proportional to its mean square error and the total number of network performance evaluations \cite{LEcuyer1994, Chan&others2023a}. 
For two unbiased estimators with the same computational cost, the relative efficiency is simply the reciprocal of the variance ratio. 
In particular, the relative efficiency of $\widehat{p}_F^{(\text{SSuR},\overline{\text{aopt}})}$ over $\widehat{p}_F^{(\text{cMCS})}$ (or $\widehat{p}_F^{(\text{MCS})}$) under the same computational cost equals: 
\begin{align}
\label{Eq: relative efficiency, practical,cMCS}
\text{relEff}_{\frac{\text{SSuR},\overline{\text{aopt}}}{\text{cMCS}}} 
&\triangleq \frac{\mathbb{V}\left( \widehat{p}_F^{(\text{cMCS})} \right)}{\mathbb{V}\left( \widehat{p}_F^{(\text{SSuR},\overline{\text{aopt}})} \right)}
= \frac{\mathbb{V}\left( \widehat{p}_F^{(\text{cMCS})} \right)}{\mathbb{V}\left( \widehat{p}_F^{(\text{SSuR},\text{opt})} \right)} \frac{1}{\alpha+1}
= \frac{1}{ r_{ \frac{\text{SSuR,opt}}{\text{cMCS}}}} \frac{1}{\alpha+1}, \\
\label{Eq: relative efficiency, practical,MCS}
\text{relEff}_{\frac{\text{SSuR},\overline{\text{aopt}}}{\text{MCS}}} 
&\triangleq \frac{\mathbb{V}\left( \widehat{p}_F^{(\text{MCS})} \right)}{\mathbb{V}\left( \widehat{p}_F^{(\text{SSuR},\overline{\text{aopt}})} \right)}
= \frac{\mathbb{V}\left( \widehat{p}_F^{(\text{MCS})} \right)}{\mathbb{V}\left( \widehat{p}_F^{(\text{SSuR},\text{opt})} \right)} \frac{1}{\alpha+1}
= \frac{1}{ r_{ \frac{\text{SSuR,opt}}{\text{MCS}}}} \frac{1}{\alpha+1}.
\end{align}
Recall that $\alpha$ is the relative increase in variance due to replacing the optimal but impractical sample size $N_{i,j}^{(\text{opt})}$ with its rounded approximation $\overline{N_{i,j}^{(\text{aopt})}}$. 
The expression of $\alpha$ is given by Eq.~(\ref{Eq: the relative variance increase due to a non-optimal sample allocation strategy}). 
The larger $\alpha$, the lower the relative efficiency. 
The variance of $\widehat{p}_F^{(\text{SSuR},\overline{\text{aopt}})}$ is estimated through independent runs of the estimator, and the variance of conditional MCS is calculated by 
$\frac{\left( \sum_{l=i^*}^n \lambda_l \right)^2 p^*_F (1 - p^*_F)}{\sum_{i,j} \overline{N_{i,j}^{(\text{aopt})}}}$. 
The variance of crude MCS is calculated by $\frac{p_F (1 - p_F)}{N^{(\text{plt})}+\sum_{i,j} \overline{N_{i,j}^{(\text{aopt})}}}$. 
$p_F$ and $p_F^*$ are taken from the ground truth. 

\subsection{DC power flow problem}
\label{Subsec: DC power flow}
We consider the DC power flow in the IEEE39 benchmark. 
Network failure is defined as the power loss (in percentage) exceeding a specified threshold, denoted as $thr$. 
The power loss is computed by solving the DC power flow problem described in Grainger \cite{Grainger1999}, and the cascading failure is modeled based on Crucitti et al. \cite{Crucitti&others2004}. 
The probabilistic inputs consist of a total of 46 Bernoulli variables, representing the state of all transmission lines, which can be either failed or safe. 
The nodes that represent the connecting buses are assumed to be in a safe state with probability one. 

\subsubsection{Ground truth}
The objectives of this subsection are threefold: (1) to provide the reference ground truth, (2) to illustrate the advantages of stratum refinement, and (3) to validate the observations discussed in Sections \ref{Sec: classic stratified sampler} and \ref{Sec: stratum refinement}. 
\paragraph{Independent and identical (IID) components}
\label{Par: Independent and identical components} 
We first consider IID inputs. 
The failure probability of each transmission line, denoted as $p$, is varied from $10^{-3}$ to $10^{-1}$, and the threshold is selected from $10\%$ to $60\%$. 
% For IID inputs, the conditional probability $p_{F \mid i}$ is decoupled with the choice of $p$. 
In the presented example, the reference conditional probability $p_{F \mid i}$ is determined through enumeration for $i \leq 5$ and $i \geq 43$, while for the remaining values of $i$, it is estimated using MCS. 
In particular, Table \ref{Table: The conditional probabilities in DC power flow problem} presents the reference conditional probabilities $p_{F \mid i}$ for $i$ from 1 to 5, and also the true minimum cut-cardinality $i^*$ for each threshold. 
Recall that according to Eq.~(\ref{Eq: p_Fi for iid inputs}), $p_{F \mid i}$ is equal to $1-\Phi_{46-i}$, so the system signature $\Phi_k$ can also be derived from Table \ref{Table: The conditional probabilities in DC power flow problem} for $k$ from 41 to 45. 

Having obtained these conditional failure probabilities, one can subsequently calculate the reference failure  probabilities, $p_F=\Pr(F)$ and $p_F^*=\Pr(F \mid I \geq i^*)$, for each scenario using Eq.~(\ref{Eq: Total probability theorem}) and Eq.~(\ref{Eq: constraint failure probability}), respectively. 
The maximum coefficient of variation (c.o.v.) of all reference failure probabilities $p_F$ is $0.023 (2.3\%)$. 
The results are listed in Table \ref{Table: The reference failure probability in DC power flow problem}, where values in parentheses present $p_F^*$ for each scenario and those outside are for $p_F$. 

We employ the stratified sampler $\widehat{p}_F^{(\text{cSS})}$ in Eq.~(\ref{Eq: conditional stratified sampler}), excluding the strata with $I < i^*$. 
Table \ref{Table: The variance reduction in DC power flow problem} shows the variance ratios, $r_{ \frac{\text{cSS,prop}}{\text{cMCS}}}$ and $r_{ \frac{\text{cSS,opt}}{\text{cMCS}}}$, for each threshold $thr$ and each component failure probability $p$. 
These ratios are computed using the reference conditional failure probabilities $p_{F \mid i}$. 
In cases where $p$ is small, even the optimal allocation strategy yields only a minor variance reduction, highlighting the necessity of conducting stratum refinement. 

\begin{table}[ht!]
    \centering
    \scriptsize
    \caption{The reference conditional probabilities $p_{F \mid i}$ and true minimum cut-cardinality $i^*$ in Example \ref{Par: Independent and identical components}.}
    \label{Table: The conditional probabilities in DC power flow problem}  
    \begin{tabular}{|c|c|c|c|c|c|c|}
	\hline
        & $i = 1$ & $i = 2$ & $i = 3$ & $i = 4$ & $i = 5$ & $i^*$\\
        \hline
        $thr=10\%$ & 0.35 & 0.60 & 0.77 & 0.87 & 0.94 & 1\\
        \hline
        $thr=20\%$ & 0.15 & 0.34 & 0.52 & 0.67 & 0.78 & 1\\
        \hline
        $thr=30\%$ & 0 & $9.2 \cdot 10^{-2}$  & 0.22 & 0.36 & 0.49 & 2\\
        \hline
        $thr=40\%$ & 0 & $1.6 \cdot 10^{-2}$  & $6.1 \cdot 10^{-2}$  & 0.13 & 0.21 & 2\\
        \hline
        $thr=50\%$ & 0 & $9.7\cdot 10^{-4}$ & $1.2 \cdot 10^{-2}$  & $3.2 \cdot 10^{-2}$  & $6.1 \cdot 10^{-2}$  & 2\\
        \hline
        $thr=60\%$ & 0 & 0 & 0 & $2.0\cdot10^{-3}$ & $6.9\cdot10^{-3}$ & 4\\
        \hline
    \end{tabular}	
\end{table}
\begin{table}[ht!]
    \centering
    \scriptsize
    \caption{The reference failure probabilities $p_F$ in Example \ref{Par: Independent and identical components}. Values in parentheses are $p_F^*$.}
    \label{Table: The reference failure probability in DC power flow problem}
    \resizebox{\textwidth}{!}{%
    \begin{tabular}{|c|c|c|c|c|c|}
	\hline
        & $p = 10^{-3}$ & $p = 5\cdot10^{-3}$ & $p = 0.01$ & $p = 0.05$ & $p = 0.1$ \\
        \hline
        $thr=10\%$ & $1.6\cdot 10^{-2}$(0.35) & $7.8\cdot 10^{-2}$(0.38) & 0.15(0.41) & 0.58(0.64) & 0.84(0.85)\\
        \hline
        $thr=20\%$ & $7.0\cdot 10^{-3}$(0.16) & $3.6\cdot 10^{-2}$(0.17) & $7.3\cdot 10^{-2}$(0.20) & 0.38(0.42) & 0.68(0.68) \\
        \hline
        $thr=30\%$ & $9.4\cdot 10^{-5}$($9.4\cdot 10^{-2}$) & $2.3\cdot 10^{-3}$(0.10) & $8.7\cdot 10^{-3}$(0.11) & 0.16(0.23) & 0.42(0.44)\\
        \hline
        $thr=40\%$ & $1.6\cdot 10^{-5}$($1.6\cdot 10^{-2}$) & $4.3\cdot 10^{-4}$($1.9\cdot 10^{-2}$) & $1.8\cdot 10^{-3}$($2.3\cdot 10^{-2}$) & $5.3\cdot 10^{-2}$($7.8\cdot 10^{-2}$) & 0.20(0.21)\\
        \hline
        $thr=50\%$ & $1.1\cdot 10^{-6}$($1.1\cdot 10^{-3}$) & $4.1\cdot 10^{-5}$($1.8\cdot 10^{-3}$) & $2.2\cdot 10^{-4}$($2.9\cdot 10^{-3}$) & $1.3\cdot 10^{-2}$($1.9\cdot 10^{-2}$) & $6.7\cdot 10^{-2}$($7.1\cdot 10^{-2}$)\\
        \hline
        $thr=60\%$ & $3.3\cdot 10^{-10}$($2.1\cdot 10^{-3}$) & $1.9\cdot 10^{-7}$($2.3\cdot 10^{-3}$) & $2.9\cdot 10^{-6}$($2.5\cdot 10^{-3}$) & $1.2\cdot 10^{-3}$($5.9\cdot 10^{-3}$) & $1.1\cdot 10^{-2}$($1.6\cdot 10^{-2}$)\\
        \hline
    \end{tabular}}	
\end{table}
\begin{table}[ht!]
    \centering
    \scriptsize
    \caption{The reference variance ratios $r_{ \frac{\text{cSS,prop}}{\text{cMCS}}}$ (shown outside parentheses) and $r_{ \frac{\text{cSS,opt}}{\text{cMCS}}}$ (shown in parentheses) in Example \ref{Par: Independent and identical components}.}
    \label{Table: The variance reduction in DC power flow problem}
    \begin{tabular}{|c|c|c|c|c|c|}
	\hline
        & $p = 10^{-3}$ & $p = 5\cdot10^{-3}$ & $p = 0.01$ & $p = 0.05$ & $p = 0.1$ \\
        \hline
        $thr=10\%$ & 0.99(0.99) & 0.97(0.97) & 0.94(0.94) & 0.82(0.79) & 0.81(0.67)\\
        \hline
        $thr=20\%$ & 0.99(0.99) & 0.97(0.96) & 0.95(0.93) & 0.82(0.81) & 0.79(0.76)\\
        \hline
        $thr=30\%$ & 1.0(0.99) & 0.99(0.97) & 0.97(0.95) & 0.87(0.83) & 0.82(0.80)\\
        \hline 
        $thr=40\%$ & 1.0(0.99) & 0.99(0.93) & 0.98(0.89) & 0.91(0.75) & 0.86(0.78)\\
        \hline
        $thr=50\%$ & 1.0(0.92) & 0.99(0.74) & 0.99(0.65) & 0.96(0.62) & 0.92(0.72)\\
        \hline
        $thr=60\%$ & 1.0(0.99) & 1.0(0.97) & 1.0(0.94) & 0.99(0.79) & 0.97(0.74)\\
        \hline
    \end{tabular}	
\end{table}

\paragraph{Independent yet non-identical (INID) components}
\label{Par: Independent yet non-identical components} 
We next consider INID components. 
In particular, we assume three different types of components. 
To this end, network edges are first arranged in decreasing order with respect to their capacities. 
The first ten edges have the failure probability $5 \cdot 10^{-3}$, while for the subsequent ten edges, the failure probability is set to $10^{-2}$. 
The probability is $0.05$ for the remaining edges. 
The threshold ranges from $10\%$ to $60\%$. 
We employ the same procedure as in Paragraph \ref{Par: Independent and identical components} to determine the reference conditional failure probabilities $p_{F \mid i}$. 
Specifically, we enumerate when $i \leq 5$ or $i \geq 43$ and perform MCS otherwise. 
The true minimum cardinality $i^*$ remains the same as in Subsection \ref{Par: Independent and identical components}. 
These results are summarized in Table \ref{Table: The conditional probabilities in DC power flow problem (NIID)}. 
Furthermore, Table \ref{Table: The reference failure probability in DC power flow problem (NIID)} shows the reference failure probabilities, $p_F$ and $p_F^*$ for each threshold setting, which can be calculated through the total probability theorem. 
The variance ratio of $\widehat{p}_F^{(\text{cSS})}$ and $\widehat{p}_F^{(\text{cMCS})}$, with either the proportional or optimal budget allocation strategy, is shown in Table \ref{Table: The variance reduction in DC power flow problem (NIID)}. 
We observe that for $thr = 10\%, 20\%$, or $30\%$, the failure probability is fairly large and crude MCS is sufficient, while for $thr = 40\%, 50\%$, and $60\%$, the variance reduction of adopting the proportional allocation strategy is negligible. 

\begin{table}[ht!]
    \centering
    \scriptsize
    \caption{The reference conditional probabilities $p_{F \mid i}$ in Example \ref{Par: Independent yet non-identical components}.} 
    \label{Table: The conditional probabilities in DC power flow problem (NIID)}
    \begin{tabular}{|c|c|c|c|c|c|c|}
	  \hline
        & $i = 1$ & $i = 2$ & $i = 3$ & $i = 4$ & $i = 5$ & $i^*$\\
        \hline
        $thr=10\%$ & 0.10 & 0.24 & 0.40 & 0.55 & 0.67 & 1\\
        \hline
        $thr=20\%$ & $3.7 \cdot 10^{-2}$ & $9.1 \cdot 10^{-2}$ & 0.17 & 0.28 & 0.38 & 1\\
        \hline
        $thr=30\%$ & 0 & $1.2 \cdot 10^{-2}$  & $3.6 \cdot 10^{-2}$  & $6.9 \cdot 10^{-2}$  & 0.11 & 2\\
        \hline
        $thr=40\%$ & 0 & $1.2 \cdot 10^{-3}$  & $3.7 \cdot 10^{-3}$  & $8.5 \cdot 10^{-3}$  & $1.6 \cdot 10^{-2}$  & 2\\
        \hline
        $thr=50\%$ & 0 & $4.5\cdot 10^{-5}$ & $4.0 \cdot 10^{-4}$  & $9.1 \cdot 10^{-4}$  & $ 1.7 \cdot 10^{-3}$  & 2\\
        \hline
        $thr=60\%$ & 0 & 0 & 0 & $1.1\cdot10^{-5}$ & $3.6\cdot10^{-5}$ & 4\\
        \hline
    \end{tabular}	
\end{table}
\begin{table}[ht!]
    \centering
    \scriptsize
    \caption{The reference failure probabilities in Example \ref{Par: Independent yet non-identical components}. Values in parentheses present $p_F^*$ for each scenario and those outside are for $p_F$. The coefficient of variation (c.o.v.) is calculated using $p_F$.}
    \label{Table: The reference failure probability in DC power flow problem (NIID)}
    \resizebox{\textwidth}{!}{%
    \begin{tabular}{|c|c|c|c|c|c|c|}
        \hline
        & $thr=10\%$ & $thr=20\%$ & $thr=30\%$ & $thr=40\%$ & $thr=50\%$ & $thr=60\%$ \\
        \hline
        $p_F(p_F^*)$ & $0.18(0.23)$ & $7.3\cdot10^{-2}(9.5\cdot10^{-2})$ & $1.2\cdot10^{-2} (2.8\cdot10^{-2})$ & $1.4\cdot10^{-3}(3.2\cdot10^{-3})$ & $1.2\cdot10^{-4}(2.9\cdot10^{-4})$ & $1.2\cdot10^{-6}(2.1\cdot10^{-5})$ \\
        \hline
        c.o.v. & $1.8\cdot10^{-5}$ & $5.1\cdot10^{-5}$ & $2.3\cdot10^{-4}$ & $8.9\cdot10^{-4}$ & $3.2\cdot10^{-3}$ & $6.5\cdot10^{-2}$ \\
        \hline
    \end{tabular}}	
\end{table}
\begin{table}[ht!]
    \centering
    \scriptsize
    \caption{The reference variance ratio $r_{ \frac{\text{cSS,prop}}{\text{cMCS}}}$ (shown outside parentheses) and $r_{ \frac{\text{cSS,opt}}{\text{cMCS}}}$ (shown in parentheses) in Example \ref{Par: Independent yet non-identical components}.}
    \label{Table: The variance reduction in DC power flow problem (NIID)}
    \resizebox{\textwidth}{!}{%
    \begin{tabular}{|c|c|c|c|c|c|c|}
	  \hline
        & $thr=10\%$ & $thr=20\%$ & $thr=30\%$ & $thr=40\%$ & $thr=50\%$ & $thr=60\%$ \\
        \hline
        variance ratio & $0.88(0.85)$ & $0.93(0.84)$ & $0.98(0.85)$ & $1.0(0.81)$ & $1.0(0.67)$ & $1.0(0.84)$ \\
        \hline
    \end{tabular}}	
\end{table}

\begin{figure}[htbp!]
    \centering
    \includegraphics[scale=0.65]{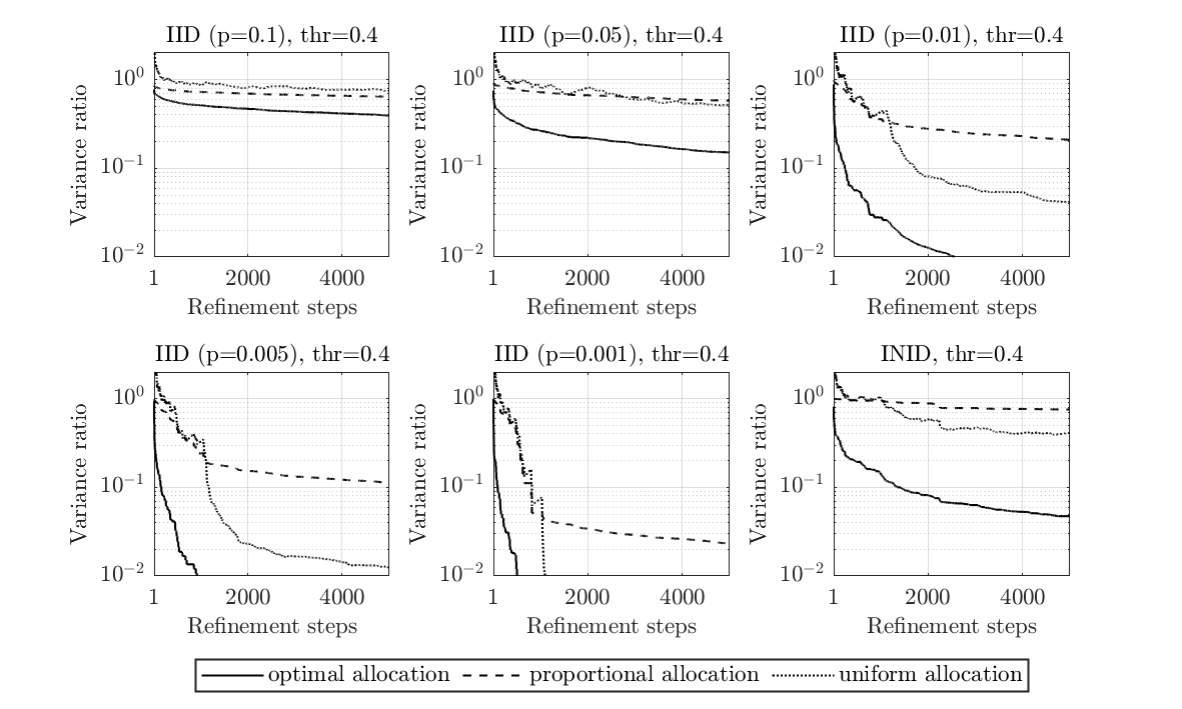}
    \caption{Variance ratios of the stratified sampler, $r_{ \frac{\text{SSuR,opt}}{\text{cMCS}}}, r_{ \frac{\text{SSuR,prop}}{\text{cMCS}}}$, and $r_{\frac{\text{SSuR,uni}}{\text{cMCS}}}$ (The threshold is $40\%$).}
    \label{Fig: Variance ratios of the stratified sampler_thr0.4.}
\end{figure}

\begin{figure}[htbp!]
    \centering
    \includegraphics[scale=0.65]{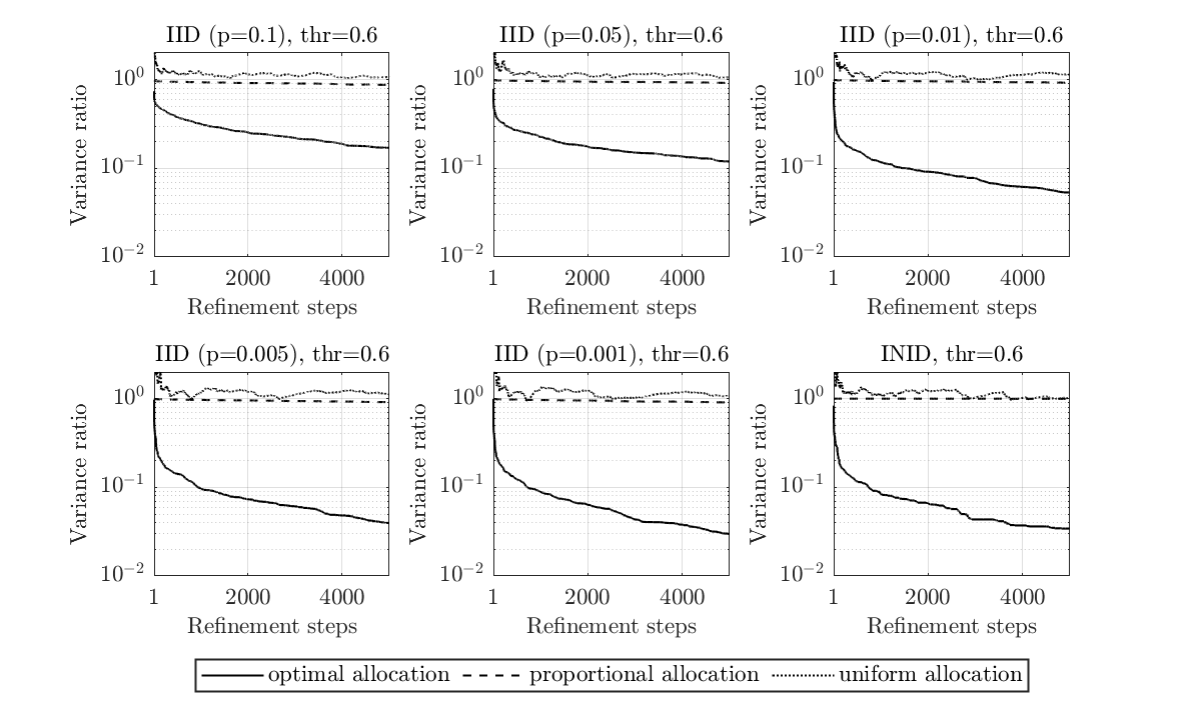}
    \caption{Variance ratios of the stratified sampler, $r_{ \frac{\text{SSuR,opt}}{\text{cMCS}}}, r_{ \frac{\text{SSuR,prop}}{\text{cMCS}}}$, and $r_{\frac{\text{SSuR,uni}}{\text{cMCS}}}$ (The threshold is $60\%$).}
    \label{Fig: Variance ratios of the stratified sampler_thr0.6.}
\end{figure}

\paragraph{Stratum refinement}
\label{Par: stratum refinement} 
In the following, we investigate how stratum refinement enhances the performance of the stratified sampler, i.e., we employ $\widehat{p}_F^{\text{(SSuR})}$ instead of $\widehat{p}_F^{\text{(cSS)}}$. 
The results are summarized in Figs.~\ref{Fig: Variance ratios of the stratified sampler_thr0.4.} and \ref{Fig: Variance ratios of the stratified sampler_thr0.6.}, where $r_{ \frac{\text{SSuR,prop}}{\text{cMCS}}}$ and $r_{ \frac{\text{SSuR,opt}}{\text{cMCS}}}$ are calculated at each step of the refinement procedure, using reference conditional failure probabilities. 
Besides the optimal and proportional strategies detailed in Section \ref{Sec: classic stratified sampler}, we consider a third strategy, termed the uniform allocation strategy, in which the computation budget is distributed uniformly among all strata. 
The corresponding variance ratio is denoted as $r_{\frac{\text{SSuR,uni}}{\text{cMCS}}}$. 
Here, we permit fractional sample sizes to enable evaluation of the variance ratios analytically, so the variance ratios do not depend on the total sample size for all three allocation strategies. 

Figs.~\ref{Fig: Variance ratios of the stratified sampler_thr0.4.} and \ref{Fig: Variance ratios of the stratified sampler_thr0.6.} show that the variance ratio of the stratified sampler using either the optimal or proportional strategies is non-increasing within each refinement step and consistently falls between zero and one. 
These align with the properties of the variance ratio discussed in Sections \ref{Sec: classic stratified sampler} and \ref{Sec: stratum refinement}. 
It is also evident from the figure that $r_{ \frac{\text{SSuR,opt}}{\text{cMCS}}}$ decreases dramatically during the initial refinement steps, demonstrating great potential, while $r_{ \frac{\text{SSuR,prop}}{\text{cMCS}}}$ decreases at a much slower rate. 
By contrast, the uniform allocation strategy can result in an estimator with a variance ratio larger than one, indicating it may be even less efficient than the conditional MCS estimator. 
Additionally, the variance ratio of the uniform allocation strategy can increase at certain refinement steps, despite the overall decreasing trend. 
We also observe that, for all three strategies, the variance ratio tends to decrease more rapidly as the component failure probability $p$ becomes smaller in the IID case. 

Recall that the reciprocal of the variance ratio indicates the relative efficiency of the stratified sampler. 
Hence, Figs.~\ref{Fig: Variance ratios of the stratified sampler_thr0.4.} and \ref{Fig: Variance ratios of the stratified sampler_thr0.6.} also depict how efficient a stratified sample can be over the conditional MCS. 
For instance, Fig.~\ref{Fig: Variance ratios of the stratified sampler_thr0.4.} indicates that the stratified sampler with proportional sample allocation is around 1.4 times more efficient than conditional MCS after around 5000 refinement steps in the 'INID' case. 
This efficiency improvement increases to 20 times if the optimal allocation strategy is available. 


\subsubsection{Numerical results} 
Next, we implement the stratified sampling summarized in Subsection \ref{Subsec: Overall workflow}. 
First, we discuss the performance of GA in identifying $i^*$. 
Then, we illustrate the efficiency of the stratified sampler, assuming $i^*$ is correctly identified. 

\paragraph{Genetic algorithm (GA) solver}
\label{Par: GA solver in Example 5.2.}
The binary-encoded GA is employed here to determine the minimum number of failed edges required to cause a system failure, i.e., $i^*$. 
The threshold $thr$ is set at 40\%, with corresponding ground truth $i^*=2$. 
For GA, we adopt the tournament selection, uniform crossover, and uniform mutation operators. 
The tournament size $n_{\text{trn}}$ and the mutation rate $p_{\text{mt}}$ are fixed at 2 and 0.01, respectively. 
The population size $n_{\text{pop}}$ ranges from 50 to 500, while the crossover fraction $f_{\text{xo}}$ varies from 0 to 1. 
We adopt the following objective function: $f_{\text{val}}(\bm{x}) = \sum_{i=1}^n x_i + (n+1)\mathbb{I}\{ \bm{x} \notin F \}$, whose minimum is the same as the solution of Eq.~(\ref{OPT: the critical number of failed components (reformulated)}). 

Because GA is a stochastic algorithm, we compute its accuracy rate, which is defined as the probability of successfully locating the global minimum, and its average computational cost over 100 independent runs. 
The results are shown in Table \ref{Table: Parameter study of the genetic algorithm}, where the values in parentheses represent the computational cost, and those outside indicate the average accuracy rates. 
It is evident that a larger population size always enhances the accuracy rate (except in cases where $f_{\text{xo}} = 0$), albeit at the expense of an increased average number of objective function calls. 
Note that GA can generate identical individuals, for which a single network performance evaluation is sufficient. 
% Also, it is observed that the application of uniform mutation with $p_{\text{mt}}=0.01$ does not enhance the overall performance. 
% For a large $p_{\text{mt}}$, we find that uniform mutation can even deteriorate the overall performance of GA. 
Table \ref{Table: GA results for cases with different thresholds} shows the accuracy rate and average computational cost for each threshold, with $f_{\text{xo}} = 0.8$ and $n_{\text{pop}} = 500$. 
Note that the accuracy rate is 0.22 for $thr=50\%$. 
This is because only one single state with two failed components can cause the system failure, making it extremely challenging for GA to identify the correct $i^*$. 
The relative biases due to misidentifying $i^*$ when $p$ is $0.001, 0.005, 0.01, 0.05, 0.1$ and in the INID case equal $-8.7\cdot10^{-4}, -0.49, -0.29, -0.02, -1.5\cdot10^{-3}$ and $-0.095$, respectively. 
In contrast, GA consistently identifies the true $i^*$ for other thresholds, and the resulting stratified sampler is unbiased. 

In the following, we specify GA parameters as: $f_{\text{xo}}=0.8, n_{\text{pop}}=500, n_{trn}=2, p_{\text{mt}} = 0.01$. 
\begin{table}[htbp!]
    \centering
    \scriptsize
    \caption{The accuracy rate (outside parentheses) and the average computational cost (in parentheses) of the genetic algorithm (GA) in Example \ref{Par: GA solver in Example 5.2.}. The tournament size $n_{\text{trn}}$ and the mutation rate $p_{\text{mt}}$ are fixed at 2 and 0.01, respectively. The population size $n_{\text{pop}}$ ranges from 50 to 500, and the crossover fraction $f_{\text{xo}}$ varies from 0 and 1.}
    \label{Table: Parameter study of the genetic algorithm}  
    \begin{tabular}{|c|c|c|c|c|}
	\hline
        & $f_{\text{xo}} = 0$ & $f_{\text{xo}} = 0.4$ & $f_{\text{xo}} = 0.8$ & $f_{\text{xo}} = 1$ \\
        \hline
        $n_{\text{pop}} = 50$ & 0.06(362) & 0.37(466) & 0.52(548) & 0.53(554) \\
        \hline
        $n_{\text{pop}} = 100$ & 0.03(778) & 0.44(985) & 0.77(1,240) & 0.93(1,308) \\
        \hline
        $n_{\text{pop}} = 200$ & 0.10(1,298) & 0.82(1,909) & 0.99(2,492) & 1(2,705) \\
        \hline
        $n_{\text{pop}} = 500$ & 0.69(3,774) & 1(5,050) & 1(6,147) & 1(6,565) \\
        \hline
    \end{tabular}	
\end{table}
\begin{table}[ht!]
    \centering
    \scriptsize
    \caption{The accuracy rate and the average computational cost of genetic algorithm (GA) in Example \ref{Par: GA solver in Example 5.2.}. The GA parameters are as follows: $n_{\text{trn}}=2$, $p_{\text{mt}}=0.01$, $n_{\text{pop}}=500$, and $f_{\text{xo}}=0.8$.}
    \label{Table: GA results for cases with different thresholds}    
    \begin{tabular}{|c|c|c|c|c|c|c|}
	  \hline
         & $thr = 10\%$ & $thr = 20\%$ & $thr = 30\%$ & $thr = 40\%$ & $thr = 50\%$ & $thr = 60\%$ \\
        \hline
        accuracy rate & 1 & 1 & 1 & 1 & 0.22 & 1\\
        \hline
        average cost &  4,009& 4,383 & 5,973 & 6,147 & 6,659 & 7,287\\
        \hline
    \end{tabular}
\end{table}

\paragraph{Approximating optimal sample allocation}
\label{Par: Approximating optimal sample allocation} 
After identifying $i^*$, we remove the redundant strata with less than $i^*$ failed components and refine the remaining strata. 
The one-step refinement procedure as described in Subsection \ref{Subsec: one-step refinement procudure} is performed 5,000 times. 
For each refined stratum, its conditional failure probability is approximated using failure individuals recorded in GA. 
The initial sample size $N$ is 10,000. 

Table \ref{Table: The reference efficiency of the stratified sampler in DC power flow problem} illustrates the relative efficiency of the proposed stratified sampler over conditional MCS, $\text{relEff}_{\frac{\text{SSuR},\overline{\text{aopt}}}{\text{cMCS}}}$, and over crude MCS, $\text{relEff}_{\frac{\text{SSuR},\overline{\text{aopt}}}{\text{MCS}}}$, estimated from 10 independent runs of the stratified sampler. 
Values for $\text{relEff}_{\frac{\text{SSuR},\overline{\text{aopt}}}{\text{MCS}}}$ are shown in parentheses and are significantly larger than those for $\text{relEff}_{\frac{\text{SSuR},\overline{\text{aopt}}}{\text{cMCS}}}$. 
In particular, the difference is related to the difference between $p_F$ and $p_F^*$ in Tables~\ref{Table: The reference failure probability in DC power flow problem} and \ref{Table: The reference failure probability in DC power flow problem (NIID)}. 
From the table, it is observed that the relative efficiency tends to decrease with increasing $p$. 
The only exception occurs when $thr = 60\%$, where the efficiency w.r.t. conditional Monte Carlo appears unaffected by changes in $p$. 
This is likely due to the poor approximation of the conditional probabilities in sample allocation. 
% There seems no monotonic trend of the relative efficiency over either the threshold $thr$ or the component failure probability $p$ from the table. 
If we assume the optimal sample allocation is available and estimate the corresponding relative efficiency, the larger the $p$ (or equivalently, the flatter the probability distribution), the lower the relative efficiency of the stratified sampler becomes for each threshold. 
The results are shown in Appendix C. 
In practice, the approximation or randomization will result in a decreased performance of the stratified sampler. 
Such influence can be quantified by the relative increase in variance, defined in Eq.~(\ref{Eq: the relative variance increase due to a non-optimal sample allocation strategy}), which is reported for each scenario in Table. \ref{Table: The relative increase in variance due to approximation and randomization of the optimal sample sizes}. 

\begin{table}[ht!]
    \centering
    \scriptsize
    \caption{The relative efficiency of the stratified sampler in Example \ref{Subsec: DC power flow}. Values outside parentheses present the relative efficiency over conditional MCS, and those inside are over crude MCS. }
    \label{Table: The reference efficiency of the stratified sampler in DC power flow problem}
    \begin{tabular}{|c|c|c|c|c|c|}
	\hline
        & IID($p = 10^{-3}$) & IID($p = 5\cdot10^{-3}$) & IID($p = 0.01$) & IID($p = 0.05$) & INID\\
        \hline
        $thr=30\%$ & $2.9\cdot10^3$($2.3\cdot10^6$) &  $1.2\cdot10^2$($4.4\cdot10^3$) & 37($3.8\cdot10^2$) & 2.7(3.1) & 3.9(6.7) \\
        \hline
        $thr=40\%$ & $1.2\cdot10^3$($8.7\cdot10^5$) &  63($2.0\cdot10^3$)& 20($1.8\cdot10^2$) & 2.5(2.6) & 3.4(5.5)\\
        \hline
        $thr=50\%$ & $3.7\cdot10^2$($2.5\cdot10^5$) &  24($7.3\cdot10^2$)& 12($1.0\cdot10^2$) & 2.3(2.3) & 2.8(4.4)\\
        \hline
        $thr=60\%$ & 0.67($3.0\cdot10^6$) & 0.72($5.8\cdot10^3$) & 0.84($5.1 \cdot 10^2$) & 1.5(5.2) & 1.2($15$)\\
        \hline
    \end{tabular}
\end{table}
\begin{table}[ht!]
    \centering
    \scriptsize
    \caption{The relative increase in variance due to approximation and randomization of the optimal sample sizes in Example \ref{Subsec: DC power flow}.}
    \label{Table: The relative increase in variance due to approximation and randomization of the optimal sample sizes}
    \begin{tabular}{|c|c|c|c|c|c|}
	\hline
        & IID($p = 10^{-3}$) & IID($p = 5\cdot10^{-3}$) & IID($p = 0.01$) & IID($p = 0.05$) & INID\\
        \hline
        $thr=30\%$ & 5.2 & 4.3 & 3.0 & 1.1 & 2.1 \\
        \hline
        $thr=40\%$ & 19 & 12 & 7.9 & 1.6 & 5.3 \\
        \hline
        $thr=50\%$ & 55 & 30 & 17 & 3.0 & 13 \\
        \hline
        $thr=60\%$ & 49 & 34 & 21 & 4.8 & 23\\
        \hline
    \end{tabular}
\end{table} 
% \begin{figure}[htbp]
%     \centering
%     \includegraphics[scale=0.7]{Relative_increase_in_variance.pdf}
%     \caption{Relative increase in variance ($thr=40\%$, IID($p=5\cdot10^{-3}$)).}
%     \label{Fig: Relative increase in variance (distributed across strata)}
% \end{figure} 

\begin{figure}[htbp]
    \centering 
    \includegraphics[scale=0.8]{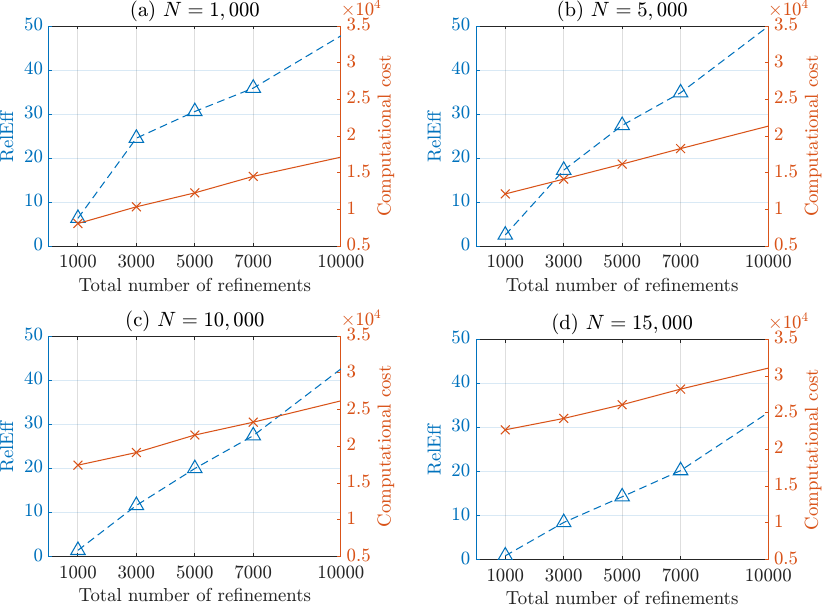}
    \caption{Influence of the number of refinement steps on $\text{relEff}_{\frac{\text{SSuR},\overline{\text{aopt}}}{\text{cMCS}}}$ and the total computational cost.} 
    \label{Fig: Influence of the number of refinement steps.} 
\end{figure}

\paragraph{The number of refinement steps}
Fig.~\ref{Fig: Influence of the number of refinement steps.} illustrates the relative efficiency and computational cost of the stratified sampler in relation to the total number of refinement steps across four different initial sample sizes, $N$. 
The threshold is $40\%$ and the probabilistic inputs are IID distributed components with failure probability $0.01$. 
In the figure, the blue dashed line shows the relative efficiency over conditional MCS, which increases with the number of refinement steps, indicating a higher efficiency of the stratified sampler. 
Note that relative efficiency is a metric that considers both the error estimate and the cost. 
% The computational cost alone also increases, as depicted by the red solid line. 
% Throughout our numerical examples, the error estimate is the mean square error computed from 10 independent runs. 
The cost is measured by the number of evaluations of the network performance function and consists of the costs for GA and the costs for the subsequent stratified sampler. 

Fig.~\ref{Fig: Influence of the number of refinement steps.} shows that, despite having the same initial sample size $N$, the actual computational cost for the stratified sampler increases with the total number of refinement steps. 
This is due to the increasing number of strata with conditional failure probabilities estimated as zero, which occurs as the strata become finer. 
According to the randomization strategy as described in Subsection \ref{Subsubsec: Randomizing the sample size}, the sample size of these strata is assigned one to guarantee an unbiased estimator, making the actual cost of the stratified sampler significantly larger than the initial sample size $N$. 
Fixing $N$, the relative efficiency of the stratified sampler also increases with the number of refinements. 
On the other hand, with a fixed number of refinements, we observe a decrease in relative efficiency as $N$ increases. 
This occurs because strata with relatively large conditional failure probabilities are inaccurately estimated as zero in the approximation. 
Consequently, the relative difference between the optimal sample size and approximated sample size (which always equals one) becomes increasingly significant as $N$ increases. 
As shown in Eq~(\ref{Eq: the relative variance increase due to a non-optimal sample allocation strategy}), the relative difference contributes quadratically to the relative increase in variance, thereby decreasing the performance of the estimator in Eq.~(\ref{Eq: relative efficiency, practical,cMCS}). 

\subsection{Source-terminal connectivity of a water supply system}
\label{Subsec: Connectivity of a water supply system}
\begin{figure}[htbp!]
    \centering
    \includegraphics[scale=0.7]{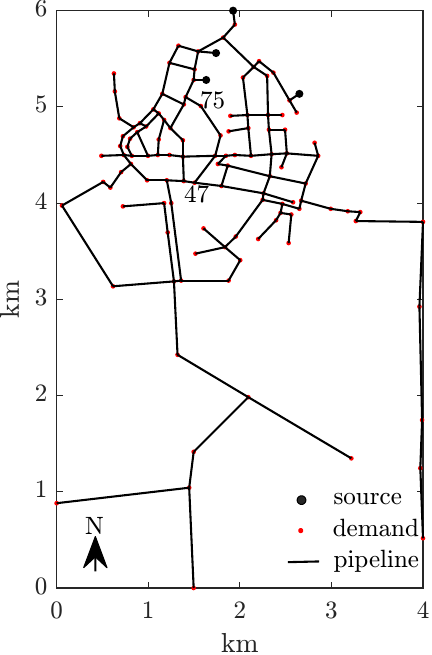}
    \caption{Topology of the water supply system in Example \ref{Subsec: Connectivity of a water supply system}.}
    \label{Fig: WaterSupplySystemTopology.}
\end{figure}

We examine the water supply system in Mianzhu, China \cite{Miao2018}. 
The network topology is mapped onto a 4 by 6-kilometer area, as illustrated in Fig.~\ref{Fig: WaterSupplySystemTopology.}. 
In this figure, lines represent pipelines and nodes represent the water sources (in black) and demands (in red). 
The system includes four source nodes, each representing a water plant in the city, and 114 demand nodes that require water.
These nodes are connected by 139 pipelines buried underground with diameters between 200mm and 500mm. 
We use the connectivity between water sources and the demands as the network performance metric and estimate the probability that a target demand node is disconnected from any of the four water plants. 
That is, if any water plant is connected to the target node, the system is safe. 
Without loss of generality, we focus on two specific demand nodes, node 47 and node 75. 
Due to lack of data, we assume that the damage of each pipeline is independent and modeled by a Poisson process along its length with a uniform failure rate, $\lambda$. 
The failure probability of the pipeline is, therefore, the probability that at least one failure occurs along the pipeline, and can be expressed as:
\begin{equation}
    p_i = 1-\exp(-\lambda l_i),    
\end{equation}
where $l_i$ is the geometric length of the $i$-th pipeline and $p_i$ denotes the failure probability of the $i$-th pipeline. 
Following Table 1 in Wang \cite{Wang1991}, the failure rate $\lambda$ is selected as either $0.1$ or $0.01$, corresponding to seismic precautionary intensities of tier seven and tier eight, respectively.  

By introducing an artificial source node that connects all four water plants with never-fail pipelines, the above reliability problem is converted into a classic two-terminal connectivity problem with INID components. 
In addition, to provide a more comprehensive analysis, we also include the results for IID distributed components, where each pipeline fails independently with the same probability $p$. 
The probability $p$ ranges from $10^{-3}$ to $0.05$. 

\subsubsection{Ground truth}
\label{Subsubsec: ground truth of water supply system example}
Table~\ref{Table: The reference failure probability of a water supply system} shows the reference failure probability $p_F$ and $p_F^*$ for each scenario. 
The latter represents the failure probability conditional on $I \geq i^*$, and is shown in parentheses in the table.  
The cardinality of the minimum cut, $i^*$, that disconnects the source from nodes 47 and 75 equals three and two, respectively.
\begin{table}[ht!]
    \centering
    \scriptsize
    \caption{The reference failure probabilities in Example \ref{Subsec: Connectivity of a water supply system}. Values inside parentheses are for $p_F^*$, the failure probability conditional on $I \geq i^*$.}
    \label{Table: The reference failure probability of a water supply system}
    \resizebox{\textwidth}{!}{%
    \begin{tabular}{|c|c|c|c|c|c|c|}
        \hline
        & IID($p=10^{-3}$) & IID($p=0.005$) & IID($p=0.01$) & IID($p=0.05$) & INID($\lambda=0.1$) & INID($\lambda=0.01$) \\
        \hline
        node 47 & $1.0\cdot10^{-9}(2.6\cdot10^{-6})$ & $1.4\cdot10^{-7}(4.2\cdot10^{-6})$ & $1.2\cdot10^{-6}(7.4\cdot10^{-6})$ & $3.1\cdot10^{-4} (3.2\cdot10^{-4})$ & $2.6\cdot10^{-5}(3.5\cdot10^{-5})$ & $1.3\cdot10^{-8}(1.8\cdot10^{-6})$ \\
        \hline
        node 75 & $2.0\cdot10^{-6}(2.3\cdot10^{-4})$ & $5.0\cdot10^{-5}(3.3\cdot10^{-4})$ & $2.0\cdot10^{-4}(5.0\cdot10^{-4})$ & $5.3\cdot10^{-3}(5.3\cdot10^{-3})$ & $1.1\cdot10^{-3}(1.2\cdot10^{-3} )$ & $1.1\cdot10^{-5}(1.9\cdot10^{-4} )$ \\
        \hline
    \end{tabular}}
\end{table} 

\begin{figure}[htbp!]
    \centering
    \includegraphics[scale=0.7]{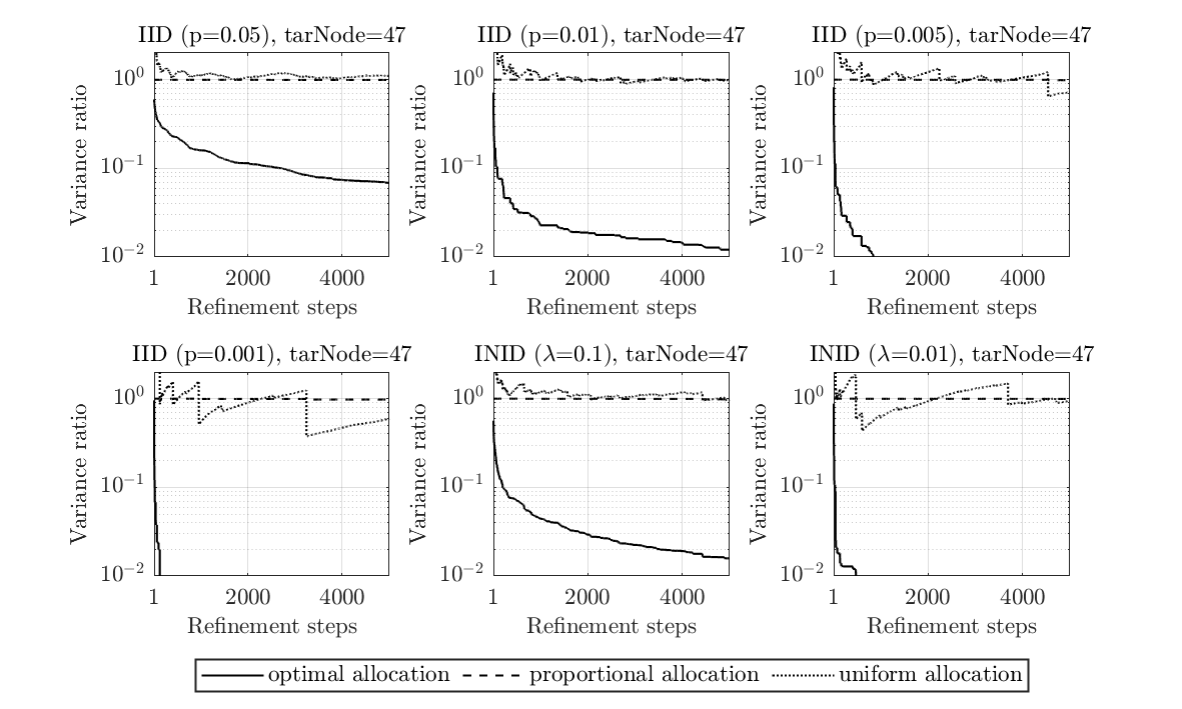}
    \caption{Variance ratios of the stratified sampler, $r_{ \frac{\text{SSuR,opt}}{\text{cMCS}}}, r_{ \frac{\text{SSuR,prop}}{\text{cMCS}}}$, and $r_{\frac{\text{SSuR,uni}}{\text{cMCS}}}$ in Example \ref{Subsec: Connectivity of a water supply system} (The target node is 47).}
    \label{Fig: Variance ratios of the stratified sampler_conn_t47.}
\end{figure}

\begin{figure}[htbp!]
    \centering
    \includegraphics[scale=0.7]{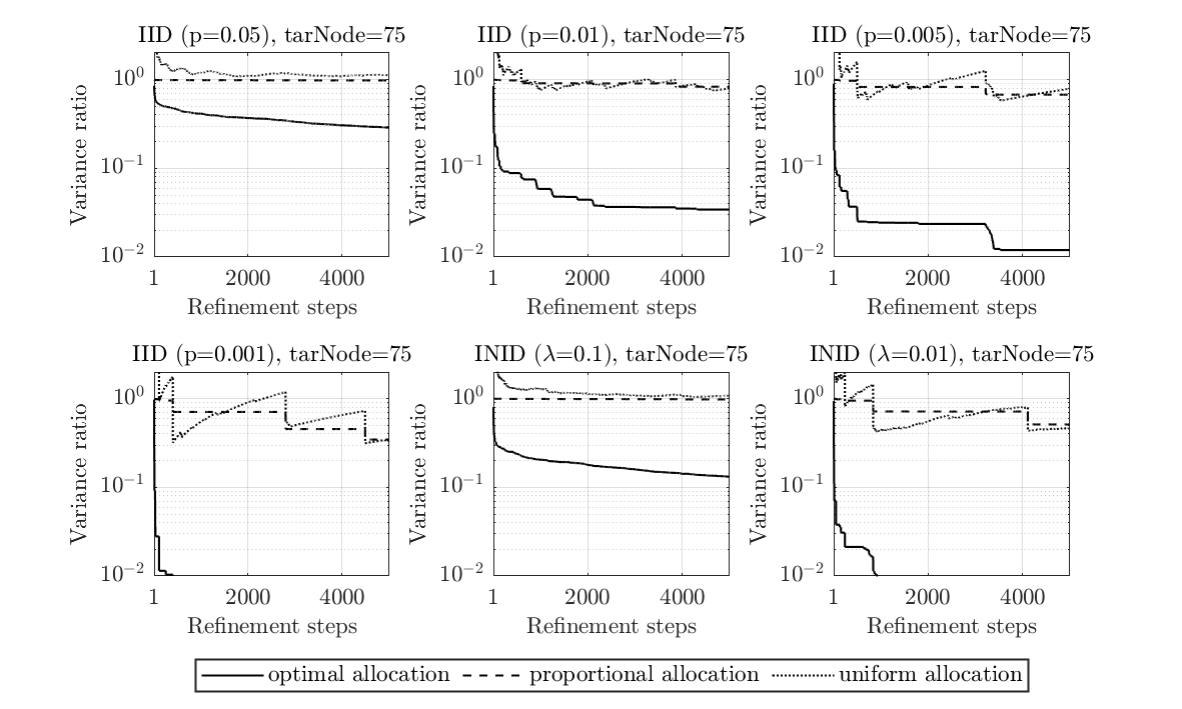}
    \caption{Variance ratios of the stratified sampler, $r_{ \frac{\text{SSuR,opt}}{\text{cMCS}}}, r_{ \frac{\text{SSuR,prop}}{\text{cMCS}}}$, and $r_{\frac{\text{SSuR,uni}}{\text{cMCS}}}$ in Example \ref{Subsec: Connectivity of a water supply system} (The target node is 75).}
    \label{Fig: Variance ratios of the stratified sampler_conn_t75.}
\end{figure}

To illustrate how the stratum refinement improves the performance of the stratified sampling, the variance ratios $r_{ \frac{\text{SSuR,opt}}{\text{cMCS}}}$ and $r_{ \frac{\text{SSuR,prop}}{\text{cMCS}}}$ are plotted against the number of refinement steps in Figs.~\ref{Fig: Variance ratios of the stratified sampler_conn_t47.} and \ref{Fig: Variance ratios of the stratified sampler_conn_t75.}. 
Both ratios decrease as the refinement iteration increases. 
In contrast, uniform sample allocation can lead to an increase in the variance ratio as the refinement progresses. 
In all cases, the stratified sampler with proportional sample allocation performs similarly to conditional MCS, showing significantly less variance reduction compared to the stratified sampler with optimal sample allocation. 
Consequently, while implementing proportional sample allocation is straightforward, the focus should be on on accurate assessment of optimal sample allocation. 


\subsubsection{Numerical results}
\label{Subsubsec: numerical results of water supply system example}
In this section, we demonstrate the results of the stratified sampler in practice, where the optimal sample size has to be approximated and subsequently randomized to an integer. 
The workflow is detailed in Subsection \ref{Subsec: Overall workflow}.

Thanks to the max-flow min-cut theorem, $i^*$ can be accurately identified through a maximum flow analysis assuming unit line capacity for each pipeline. 
Next, we perform 5,000 iterations of stratum refinement and estimate the conditional failure probability for each stratum using minimal cuts and the coherency property of the network performance function. 
Obvious choices of the minimal cut set include, for example, (1) the three (or two) pipelines directly connected to node 47 (or 75) and (2) the four pipelines directly connected to one of the four water plants. 
Finally, we perform the stratified sampler with an initial sample size $N$ of 10,000. 
The optimal sample size per stratum is computed using the estimated conditional failure probabilities and is subsequently randomized into a neighboring integer. 

The relative efficiency of the final stratified sampler is illustrated in Table \ref{Table: The reference efficiency of the stratified sampler in water supply system example} after 10 independent runs of the stratified sampler. 
For IID distributed components, we observe a decrease in relative efficiency, both $\text{relEff}_{\text{opt,cMCS}}$ and $\text{relEff}_{\text{prop,MCS}}$, when the component failure probability $p$ increases. 
A similar observation can be made for INID distributed components. 
In addition, the larger the ratio $p^*_F / p_F$ in Table \ref{Table: The reference failure probability of a water supply system}, the larger the ratio $r_{ \frac{\text{SSuR,opt}}{\text{cMCS}}}\big/r_{ \frac{\text{SSuR,opt}}{\text{MCS}}}$, and consequently, the greater the increase in relative efficiency when redundant strata are removed. 
Due to the approximation and randomization of the optimal sample sizes, the variance of the stratified sampler is significantly larger than the minimum variance under the same computational cost, i.e., the number of network performance function evaluations. 
The relative increase in variance, as defined in Eq.~(\ref{Eq: the relative variance increase due to a non-optimal sample allocation strategy}), is reported for each scenario in Table. \ref{Table: The relative increase in variance due to approximation and randomization of the optimal sample sizes in water supply system example}. 
To further improve the stratified sampler's performance, more minimal cuts can be selected to form a better approximation of the optimal sample size. 

\begin{table}[ht!]
    \centering
    \scriptsize
    \caption{The relative efficiency of the stratified sampler in Example \ref{Subsec: Connectivity of a water supply system}. Values in parentheses present the relative efficiency over crude MCS, and those outside are over conditional MCS.} 
    \label{Table: The reference efficiency of the stratified sampler in water supply system example}
    \resizebox{\textwidth}{!}{%
    \begin{tabular}{|c|c|c|c|c|c|c|}
	\hline
        & IID($p = 10^{-3}$) & IID($p = 5\cdot10^{-3}$) & IID($p = 0.01$) & IID($p = 0.05$) & INID($\lambda=0.1$) & INID($\lambda=0.01$)\\
        \hline
        node 47 & $1.4\cdot10^2$($3.5\cdot10^5$) &  19($5.7\cdot10^2$) & 14(87) & 2.2(2.2) & 6.7(8.9) & 44($5.9\cdot10^3$)\\
        \hline
        node 75 & $4.7\cdot10^2$($5.3\cdot10^4$) &  55($3.6\cdot10^2$)& 19(47) & 1.8(1.8) & 3.9(4.3) & $1.6\cdot10^2$($2.7\cdot10^3$)\\
        \hline
    \end{tabular}}
\end{table}
\begin{table}[ht!]
    \centering
    \scriptsize
    \caption{The relative increase in variance due to approximation and randomization of the optimal sample sizes in Example \ref{Subsec: Connectivity of a water supply system}.}
    \label{Table: The relative increase in variance due to approximation and randomization of the optimal sample sizes in water supply system example}
    \resizebox{\textwidth}{!}{%
    \begin{tabular}{|c|c|c|c|c|c|c|}
	\hline
        & IID($p = 10^{-3}$) & IID($p = 5\cdot10^{-3}$) & IID($p = 0.01$) & IID($p = 0.05$) & INID($\lambda=0.1$) & INID($\lambda=0.01$)\\
        \hline
        node 47 & 12 & 12 & 4.8 & 5.8 & 8.5 & 14\\
        \hline
        node 75 & 0.52 & 0.53 & 0.54 & 0.92 & 0.94 & 0.53\\
        \hline
    \end{tabular}}
\end{table}

\section{Concluding remarks}
The main contribution of this work is a novel stratified sampler with unbalanced stratum refinement, specifically designed for network reliability assessment. 
The rationale behind this approach lies in the observation that the variance ratio of the stratified sampler with either proportional or optimal sample allocation will not increase after refining any stratum. 
Each stratum is defined by a partition of network components and a specific configuration of failed components. 
We utilize the conditional Bernoulli model to sample within each stratum and to calculate the stratum's size or probability. 
By removing redundant strata where the number of failed components in any state is below the minimum $i^*$ required for failure to occur, the performance of the stratified sampler can be further enhanced. 
The resulting variance reduction is significant when a large portion of the probability is removed. 
We discuss strategies for identifying $i^*$ for connectivity problems and employ the genetic algorithm to estimate $i^*$ for problems with physics-based performance metrics.
We find that the stratified sampler with proportional sample allocation often leads to a marginal variance reduction compared to conditional Monte Carlo simulation given failure of at least $i^*$ components. 
Consequently, we propose a heuristic for approximating optimal sample allocation, which is subsequently combined with a randomization strategy to ensure an integer sample size within each stratum and, hence, an unbiased estimator of the failure probability. 
% The heuristic is based on the minimal cuts or the failed individuals in GA and assumes a coherent network performance function. 

Across all scenarios in our numerical examples the stratified sampler outperforms clearly both crude and conditional Monte Carlo, with the expectation of case $thr=60\%$ in Example 1, where the sampler is slightly worse than conditional Monte Carlo. 
This is attributed to the poor approximation and randomization of the optimal sample size. 
We also found that the flatter the probability distribution, the lower the relative efficiency of the stratified sampler becomes. 
 
We note that the proposed stratified sampler is tailored to independent and binary inputs. 
In practice, components can have multiple failure states and can fail dependently. 
The extension of the refined stratified sampler to incorporate the common cause failure or multi-state components is deferred to future work. 
Additionally, the primary motivation behind the system signature is to decouple system reliability computation from component probabilities, enabling efficient reliability evaluation across different time points. In contrast, the proposed method optimizes the strata based on fixed component reliabilities. However, if component reliabilities vary over time, a natural question arises: Which reliability values should be used to define the strata? This presents a promising direction for future research.

\section{Acknowledgement}
We thank Prof.Ji-Eun Byun at the University of Glasgow for her insightful discussion. 

\appendix

\section{Upper and lower bounds of $r_{ \frac{\text{SS,prop}}{\text{MCS}}}$ and $r_{ \frac{\text{SS,opt}}{\text{MCS}}}$}
In this section, we prove the upper and lower bounds of the variance ratios $r_{ \frac{\text{SS,prop}}{\text{MCS}}}$ and $r_{ \frac{\text{SS,opt}}{\text{MCS}}}$ defined by Eqs.(\ref{Eq: Variance ratio 1}) and (\ref{Eq: Variance ratio 2}), respectively. 
The notation in this section is the same as in Section \ref{Sec: classic stratified sampler}. 

The key to the derivation of the bounds is to treat the serial number of stratum, i.e., $i$, as an allocation variable, denoted as $I$, with a probability distribution given by $\Pr(I=i)=\lambda_i$. 
Consequently, it holds that:
\begin{equation}
\label{Eq: variance of the conditional failure probability}
\sum_{i=0}^n \lambda_i p^2_{F\mid i} - \left( \sum_{i=0}^n \lambda_i p_{F\mid i} \right)^2 
= \mathbb{V}_I \left( p_{F\mid I} \right)
= \mathbb{V}_I \left[ \mathbb{E}_{\bm{X} \mid I} \left( \mathbb{I}\{\bm{X} \in F\} \right) \right]  
\geq 0.
\end{equation}
By using Eq.~(\ref{Eq: variance of the conditional failure probability}), one can derive:
\begin{equation}
\label{Eq: upper bound of r_prop}
r_{ \frac{\text{SS,prop}}{\text{MCS}}} 
= 1 - \frac{ \mathbb{V}_I\left[\mathbb{E}_{\bm{X} \mid I} \left( \mathbb{I}\{\bm{X} \in F\} \right) \right] }{p_F - p_F^2 } \leq 1. 
\end{equation}
Equality in (\ref{Eq: upper bound of r_prop}) is achieved if and only if $\mathbb{V}_I\left[\mathbb{E}_{\bm{X} \mid I} \left( \mathbb{I}_F\{\bm{X}\} \right) \right] = 0$. 
In other words, if the conditional failure probability $p_{F \mid i} \triangleq \mathbb{E}_{\bm{X} \mid I=i} \left( \mathbb{I}_F\{\bm{X}\} \right)$ is invariant over different strata, and equals the failure probability $p_F$. 
As for the optimal variance ratio $r_{ \frac{\text{SS,opt}}{\text{MCS}}}$, since it is not larger than $r_{ \frac{\text{SS,prop}}{\text{MCS}}}$, it holds that $r_{ \frac{\text{SS,opt}}{\text{MCS}}} = 1 \Rightarrow r_{ \frac{\text{SS,prop}}{\text{MCS}}} = 1 \Rightarrow \forall i:p_{F \mid i } = p_F$. 
By observing Eq.~(\ref{Eq: Variance ratio 2}), it is evident that $\forall i:p_{F \mid i } = p_F \Rightarrow r_{ \frac{\text{SS,opt}}{\text{MCS}}} = 1$. 

\iffalse
By further noting that $\mathbb{V}_{\bm{X}}\left( \mathbb{I}\{\bm{X}\in F\} \right) = p_F - p_F^2$, and by using the variance decomposition formula, one obtains:
\begin{align}
r_{ \frac{\text{SS,prop}}{\text{MCS}}} 
&= \frac{ \mathbb{V}_{\bm{X}}\left( \mathbb{I}\{\bm{X}\in F\} \right) - \mathbb{V}_I\left[\mathbb{E}_{\bm{X} \mid I} \left( \mathbb{I}\{\bm{X}\in F\} \right)\right] }{ \mathbb{V}_{\bm{X}}\left( \mathbb{I} \{\bm{X}\in F\} \right) }  \notag \\
\label{Eq: lower bound of r_prop}
&= \frac{\mathbb{E}_I \left[ \mathbb{V}_{\bm{X} \mid I}(\mathbb{I}(\bm{X}) \in F) \right]}{\mathbb{V}_{\bm{X}}\left( \mathbb{I}\{\bm{X}\in F\} \right)}  \geq 0.
\end{align} 
The equality of (\ref{Eq: lower bound of r_prop}) is achieved if and only if $\mathbb{E}_I \left[ \mathbb{V}_{\bm{X} \mid I}(\mathbb{I}(\bm{X}) \in F) \right] = 0$, i.e., $\mathbb{V}_{\bm{X} \mid I}(\mathbb{I}(\bm{X}) \in F) = p_{F\mid i}(1-p_{F\mid i}) = 0$ for each stratum. 
\fi

As for the lower bounds, it is also evident that $r_{ \frac{\text{SS,prop}}{\text{MCS}}} = 0 \Leftrightarrow  r_{ \frac{\text{SS,opt}}{\text{MCS}}} = 0 \Leftrightarrow  \forall i: p_{F \mid i }(1-p_{F \mid i })=0$ by observing Eqs.~(\ref{Eq: Variance ratio 1}) and (\ref{Eq: Variance ratio 2}), since $\lambda_i > 0$ for each $i$. 

\section{A proof of the mononicity of $r_{ \frac{\text{SS,prop}}{\text{MCS}}}$ and $r_{ \frac{\text{SS,opt}}{\text{MCS}}}$ when refining the strata}

In the following, we give a proof of inequalities (\ref{Eq: rationale of stratum refinement 1}) and (\ref{Eq: rationale of stratum refinement 2}). 
Let $\mathcal{S}_{i_1}$ and $\mathcal{S}_{i_2}$ denote the two sub-strata split from the stratum $\mathcal{S}_i$. 
The probabilities of strata are denoted as $\lambda_{i_1}, \lambda_{i_2}$, and $\lambda_i$, respectively. 
The associated conditional failure probabilities are denoted as $p_{F \mid i_1}, p_{F \mid i_2}$, and $p_{F \mid i}$, respectively. 

Noting that $\lambda_{i_1}p_{F \mid i_1} + \lambda_{i_2}p_{F \mid i_2} = \lambda_ip_{F \mid i}$, proving inequality (\ref{Eq: rationale of stratum refinement 1}) is equivalent to proving:
\begin{equation}
\label{AppendixB: Inequality1}
     \lambda_{i_1} \cdot p^2_{F\mid i_1} + \lambda_{i_2} \cdot p^2_{F\mid i_2} - \lambda_i \cdot p^2_{F\mid i} \geq 0. 
\end{equation}
Multiplying both sides of Inequality (\ref{AppendixB: Inequality1}) by $\lambda_{i_1}\lambda_{i_2}\lambda_i$ and letting
\begin{equation}
a \triangleq \lambda_{i_1} p_{F\mid i_1} \leq \lambda_{i_1}, \quad b \triangleq \lambda_{i_2}  p_{F\mid i_2} \leq \lambda_{i_2},
\end{equation}
the inequality can be rewritten as:
\begin{equation}
\label{AppendixB: Inequality2}
     \lambda_{i_2} \lambda_i a^2 + \lambda_{i_1} \lambda_i b^2 - \lambda_{i_1} \lambda_{i_2} (a+b)^2 \geq 0. 
\end{equation}
The left-hand side of Inequality (\ref{AppendixB: Inequality2}) can be written as a square term, thereby concluding the poof. 
Indeed, since $\lambda_i = \lambda_{i_1} + \lambda_{i_2}$, it holds that: 
\begin{align}
\lambda_{i_2} \lambda_i a^2 & + \lambda_{i_1} \lambda_i b^2 - \lambda_{i_1} \lambda_{i_2} (a+b)^2 = (\lambda_{i_2})^2 a^2 + (\lambda_{i_1})^2 b^2 - 2\lambda_{i_1} \lambda_{i_2} ab \notag \\
& = (\lambda_{i_2}a - \lambda_{i_1}b )^2  = (\lambda_{i_1}\lambda_{i_2})^2(p_{F\mid i_1} -p _{F\mid i_2})^2 \geq 0. \notag
\end{align}
This means, if and only if $p_{F\mid i_1} =p _{F\mid i_2}$, Inequality (\ref{AppendixB: Inequality2}), and consequently Inequality (\ref{Eq: rationale of stratum refinement 1}), take the equal sign. 
In other cases, the stratum refinement leads to a reduced variance ratio $r_{ \frac{\text{SSuR,prop}}{\text{cMCS}}}$. 

Next, we prove Inequality (\ref{Eq: rationale of stratum refinement 2}). 
The problem can be reformulated as follows: 
\begin{align}
&\text{Inequality (\ref{Eq: rationale of stratum refinement 2})} \notag \\
&\Leftrightarrow  \sqrt{a(\lambda_{i_1} - a)} + \sqrt{b(\lambda_{i_2} - b)} \leq \sqrt{(a+b)(\lambda_i - a-b)} \notag \\
& \Leftrightarrow a(\lambda_{i_1} - a) + b(\lambda_{i_2} - b) + 2\sqrt{ab(\lambda_{i_1}-a)(\lambda_{i_2}-b)} \leq (a+b)(\lambda_i - a-b) \notag \\
& \Leftrightarrow 2\sqrt{ab(\lambda_{i_1}-a)(\lambda_{i_2}-b)} \leq a\lambda_{i_2} + b\lambda_{i_1} - 2ab \notag \\
& \Leftrightarrow 2\sqrt{ab(\lambda_{i_1}-a)(\lambda_{i_2}-b)} \leq a(\lambda_{i_2}-b) + b(\lambda_{i_1} - a) \notag \\
\label{AppendixB: Inequality3}
&  \Leftrightarrow \left( \sqrt{a(\lambda_{i_2}-b)} - \sqrt{b(\lambda_{i_1}-a)} \right)^2 \geq 0.
\end{align}
Clearly, a square term will be no less than zero, which concludes the proof. 
In addition, Inequality (\ref{AppendixB: Inequality3}) takes the equal sign if and only if $a(\lambda_{i_2}-b)=b(\lambda_{i_1}-a)$, or equivalently, $p_{F\mid i_1} = p_{F\mid i_2}$. 
In other cases, the stratum refinement results in a reduced variance ratio $r_{ \frac{\text{SSuR,opt}}{\text{cMCS}}}$. 

\section{Other supplementary material}
\begin{table}[htbp!]
    \centering
    \scriptsize
    \caption{The relative efficiency of the stratified sampler with optimal sample allocation, $\text{RelEff}_{\text{opt,cMCS}}$, in Example \ref{Subsec: DC power flow}. Values outside parentheses present the variance ratio over conditional MCS, and those inside are over crude MCS. }
    
    \begin{tabular}{|c|c|c|c|c|c|}
	\hline
        & IID($p = 10^{-3}$) & IID($p = 5\cdot10^{-3}$) & IID($p = 0.01$) & IID($p = 0.05$) & INID\\
        \hline
        $thr=30\%$ & $1.8\cdot10^4$($1.4\cdot10^7$) &  $6.5\cdot10^2$($2.3\cdot10^4$) & $1.5\cdot10^2$($1.5\cdot10^3$) & 5.6(6.4) & 12(21) \\
        \hline
        $thr=40\%$ & $2.4\cdot10^4$($1.7\cdot10^7$) &  $8.0\cdot10^2$($2.6\cdot10^4$)& $1.8\cdot10^2$($1.6\cdot10^3$) & 6.6(7.0) & 21(35)\\
        \hline
        $thr=50\%$ & $2.1\cdot10^4$($1.4\cdot10^7$) &  $7.6\cdot10^2$($2.3\cdot10^4$)& $2.1\cdot10^2$($1.9\cdot10^3$) & 9.0(9.2) & 39(62)\\
        \hline
        $thr=60\%$ & 34($1.5\cdot10^8$) & 25($2.0\cdot10^4$) & 19($1.1 \cdot 10^4$) & 8.5(30) & 29($3.7\cdot10^2$)\\
        \hline
    \end{tabular}
\end{table}

\iffalse
The inequality \ref{Eq: The variance of the classic stratified sampling estimator (proportional allocation)} results from the fact that 
\fi

%---------------------------------------------------------------------------------
\newpage
\bibliographystyle{IeeeTran}
\bibliography{mybib}

\begin{thebibliography}{10}
\providecommand{\url}[1]{#1}
\csname url@samestyle\endcsname
\providecommand{\newblock}{\relax}
\providecommand{\bibinfo}[2]{#2}
\providecommand{\BIBentrySTDinterwordspacing}{\spaceskip=0pt\relax}
\providecommand{\BIBentryALTinterwordstretchfactor}{4}
\providecommand{\BIBentryALTinterwordspacing}{\spaceskip=\fontdimen2\font plus
\BIBentryALTinterwordstretchfactor\fontdimen3\font minus
  \fontdimen4\font\relax}
\providecommand{\BIBforeignlanguage}[2]{{%
\expandafter\ifx\csname l@#1\endcsname\relax
\typeout{** WARNING: IEEEtran.bst: No hyphenation pattern has been}%
\typeout{** loaded for the language `#1'. Using the pattern for}%
\typeout{** the default language instead.}%
\else
\language=\csname l@#1\endcsname
\fi
#2}}
\providecommand{\BIBdecl}{\relax}
\BIBdecl

\bibitem{Li&Liu2021}
J.~Li and W.~Liu, \emph{Lifeline Engineering Systems: Network Reliability
  Analysis and Aseismic Design}.\hskip 1em plus 0.5em minus 0.4em\relax
  Springer Nature, 2021.

\bibitem{Billinton&Li1994}
R.~Billinton and W.~Li, \emph{Reliability assessment of electric power systems
  using Monte Carlo methods}.\hskip 1em plus 0.5em minus 0.4em\relax Springer
  Science \& Business Media, 1994.

\bibitem{Provan&Ball1984}
J.~S. Provan and M.~O. Ball, ``Computing network reliability in time polynomial
  in the number of cuts,'' \emph{Operations Research}, vol.~32, no.~3, pp.
  516--526, 1984.

\bibitem{Ball1986}
M.~O. Ball, ``Computational complexity of network reliability analysis: {A}n
  overview,'' \emph{IEEE Transactions on Reliability}, vol.~35, no.~3, pp.
  230--239, 1986.

\bibitem{Jane&others1993}
C.-C. Jane, J.-S. Lin, and J.~Yuan, ``Reliability evaluation of a limited-flow
  network in terms of minimal cutsets,'' \emph{IEEE Transactions on
  Reliability}, vol.~42, no.~3, pp. 354--361, 1993.

\bibitem{Zuo&others2007}
M.~J. Zuo, Z.~Tian, and H.-Z. Huang, ``An efficient method for reliability
  evaluation of multistate networks given all minimal path vectors,'' \emph{IIE
  Transactions}, vol.~39, no.~8, pp. 811--817, 2007.

\bibitem{Brown&others2021}
J.~I. Brown, C.~J. Colbourn, D.~Cox, C.~Graves, and L.~Mol, ``Network
  reliability: {H}eading out on the highway,'' \emph{Networks}, vol.~77, no.~1,
  pp. 146--160, 2021.

\bibitem{Imai&others1999}
H.~Imai, K.~Sekine, and K.~Imai, ``Computational investigations of all-terminal
  network reliability via {B}{D}{D}s,'' \emph{IEICE Transactions on
  Fundamentals of Electronics, Communications and Computer Sciences}, vol.~82,
  no.~5, pp. 714--721, 1999.

\bibitem{Hardy&others2007}
G.~Hardy, C.~Lucet, and N.~Limnios, ``K-terminal network reliability measures
  with binary decision diagrams,'' \emph{IEEE Transactions on Reliability},
  vol.~56, no.~3, pp. 506--515, 2007.

\bibitem{Levitin&others2003}
G.~Levitin, L.~Podofillini, and E.~Zio, ``Generalised importance measures for
  multi-state elements based on performance level restrictions,''
  \emph{Reliability Engineering \& System Safety}, vol.~82, no.~3, pp.
  287--298, 2003.

\bibitem{Li&Zio2012}
Y.-F. Li and E.~Zio, ``A multi-state model for the reliability assessment of a
  distributed generation system via universal generating function,''
  \emph{Reliability Engineering \& System Safety}, vol. 106, pp. 28--36, 2012.

\bibitem{Song&Kang2009}
J.~Song and W.-H. Kang, ``System reliability and sensitivity under statistical
  dependence by matrix-based system reliability method,'' \emph{Structural
  Safety}, vol.~31, no.~2, pp. 148--156, 2009.

\bibitem{Li&He2002}
J.~Li and J.~He, ``A recursive decomposition algorithm for network seismic
  reliability evaluation,'' \emph{Earthquake Engineering \& Structural
  Dynamics}, vol.~31, no.~8, pp. 1525--1539, 2002.

\bibitem{Lim&Song2012}
H.-W. Lim and J.~Song, ``Efficient risk assessment of lifeline networks under
  spatially correlated ground motions using selective recursive decomposition
  algorithm,'' \emph{Earthquake Engineering \& Structural Dynamics}, vol.~41,
  no.~13, pp. 1861--1882, 2012.

\bibitem{Paredes&others2018}
R.~Paredes, L.~Due{\~n}as-Osorio, and I.~Hernandez-Fajardo, ``Decomposition
  algorithms for system reliability estimation with applications to
  interdependent lifeline networks,'' \emph{Earthquake Engineering \&
  Structural Dynamics}, vol.~47, no.~13, pp. 2581--2600, 2018.

\bibitem{Byun&others2024}
J.-E. Byun, H.~Ryu, and D.~Straub, ``Branch-and-bound algorithm for efficient
  reliability analysis of general coherent systems,'' \emph{arXiv preprint
  arXiv:2410.22363}, 2024.

\bibitem{Zio&Pedroni2008}
E.~Zio and N.~Pedroni, ``Reliability analysis of discrete multi-state systems
  by means of subset simulation,'' in \emph{Proceedings of the 17th ESREL
  Conference}, Valencia, Spain, 2008, pp. 22--25.

\bibitem{Zuev&others2015}
K.~M. Zuev, S.~Wu, and J.~L. Beck, ``General network reliability problem and
  its efficient solution by subset simulation,'' \emph{Probabilistic
  Engineering Mechanics}, vol.~40, pp. 25--35, 2015.

\bibitem{Jensen&Jerez2018}
H.~A. Jensen and D.~J. Jerez, ``A stochastic framework for reliability and
  sensitivity analysis of large scale water distribution networks,''
  \emph{Reliability Engineering \& System Safety}, vol. 176, pp. 80--92, 2018.

\bibitem{Chan&others2022a}
J.~Chan, I.~Papaioannou, and D.~Straub, ``An adaptive subset simulation
  algorithm for system reliability analysis with discontinuous limit states,''
  \emph{Reliability Engineering \& System Safety}, vol. 225, p. 108607, 2022.

\bibitem{Hui&others2005}
K.-P. Hui, N.~Bean, M.~Kraetzl, and D.~P. Kroese, ``The cross-entropy method
  for network reliability estimation,'' \emph{Annals of Operations Research},
  vol. 134, no.~1, pp. 101--118, 2005.

\bibitem{Chan&others2023a}
J.~Chan, I.~Papaioannou, and D.~Straub, ``Bayesian improved cross entropy
  method for network reliability assessment,'' \emph{Structural Safety}, vol.
  103, p. 102344, 2023.

\bibitem{Chan&others2024}
------, ``Bayesian improved cross entropy method with categorical mixture
  models,'' \emph{Reliability Engineering \& System Safety}, vol. 252, p.
  110432, 2024.

\bibitem{VanSlyke&Frank1971}
R.~Van~Slyke and H.~Frank, ``Network reliability analysis: {P}art 1,''
  \emph{Networks}, vol.~1, no.~3, pp. 279--290, 1971.

\bibitem{Chan&others2025}
J.~Chan, R.~Paredes, I.~Papaioannou, L.~Duenas-Osorio, and D.~Straub,
  ``Adaptive monte carlo methods for estimating rare events in power grids,''
  \emph{ASCE-ASME Journal of Risk and Uncertainty in Engineering Systems, Part
  A: Civil Engineering}, vol.~11, no.~1, p. 04024082, 2025.

\bibitem{Duenas&others2017}
L.~Duenas-Osorio, K.~Meel, R.~Paredes, and M.~Vardi, ``Counting-based
  reliability estimation for power-transmission grids,'' in \emph{Proceedings
  of the AAAI Conference on Artificial Intelligence}, vol.~31, no.~1, San
  Francisco, California, USA, 2017.

\bibitem{Elperin&others1991}
T.~Elperin, I.~Gertsbakh, and M.~Lomonosov, ``Estimation of network reliability
  using graph evolution models,'' \emph{IEEE Transactions on Reliability},
  vol.~40, no.~5, pp. 572--581, 1991.

\bibitem{Cancela&others2022}
H.~Cancela, L.~Murray, and G.~Rubino, ``Reliability estimation for stochastic
  flow networks with dependent arcs,'' \emph{IEEE Transactions on Reliability},
  pp. 622--636, 2022.

\bibitem{Cancela&Khadiri1995}
H.~Cancela and M.~El~Khadiri, ``A recursive variance-reduction algorithm for
  estimating communication-network reliability,'' \emph{IEEE Transactions on
  Reliability}, vol.~44, no.~4, pp. 595--602, 1995.

\bibitem{Dehghani&others2021}
N.~L. Dehghani, S.~Zamanian, and A.~Shafieezadeh, ``Adaptive network
  reliability analysis: {M}ethodology and applications to power grid,''
  \emph{Reliability Engineering \& System Safety}, vol. 216, p. 107973, 2021.

\bibitem{Ding&others2024}
C.~Ding, P.~Wei, Y.~Shi, J.~Liu, M.~Broggi, and M.~Beer, ``Sampling and active
  learning methods for network reliability estimation using {K}-terminal
  spanning tree,'' \emph{Reliability Engineering \& System Safety}, vol. 250,
  p. 110309, 2024.

\bibitem{Coolen&Coolen-Maturi2024}
F.~P. Coolen and T.~Coolen-Maturi, ``Survival signature for reliability
  quantification of large systems and networks,'' in \emph{System Dependability
  - Theory and Applications}, W.~Zamojski, J.~Mazurkiewicz, J.~Sugier,
  T.~Walkowiak, and J.~Kacprzyk, Eds.\hskip 1em plus 0.5em minus 0.4em\relax
  Springer Nature Switzerland, 2024, pp. 29--37.

\bibitem{DiMaio&others2023}
F.~Di~Maio, C.~Pettorossi, and E.~Zio, ``Entropy-driven {M}onte {C}arlo
  simulation method for approximating the survival signature of complex
  infrastructures,'' \emph{Reliability Engineering \& System Safety}, vol. 231,
  p. 108982, 2023.

\bibitem{Tong2006}
C.~Tong, ``Refinement strategies for stratified sampling methods,''
  \emph{Reliability Engineering \& System Safety}, vol.~91, no. 10-11, pp.
  1257--1265, 2006.

\bibitem{Shields&others2015}
M.~D. Shields, K.~Teferra, A.~Hapij, and R.~P. Daddazio, ``Refined stratified
  sampling for efficient {M}onte {C}arlo based uncertainty quantification,''
  \emph{Reliability Engineering \& System Safety}, vol. 142, pp. 310--325,
  2015.

\bibitem{Cochran1977}
W.~G. Cochran, \emph{Sampling Techniques}.\hskip 1em plus 0.5em minus
  0.4em\relax John Wiley \& Sons, 1977.

\bibitem{Song&Kawai2023}
C.~Song and R.~Kawai, ``Adaptive stratified sampling for structural reliability
  analysis,'' \emph{Structural Safety}, vol. 101, p. 102292, 2023.

\bibitem{Etore&others2011}
P.~Etor{\'e}, G.~Fort, B.~Jourdain, and E.~Moulines, ``On adaptive
  stratification,'' \emph{Annals of operations research}, vol. 189, pp.
  127--154, 2011.

\bibitem{Pettersson&Krumscheid2022}
P.~Pettersson and S.~Krumscheid, ``Adaptive stratified sampling for nonsmooth
  problems,'' \emph{International Journal for Uncertainty Quantification},
  vol.~12, no.~6, pp. 71--99, 2022.

\bibitem{Fishman1989}
G.~S. Fishman, ``{M}onte {C}arlo estimation of the maximal flow distribution
  with discrete stochastic arc capacity levels,'' \emph{Naval Research
  Logistics (NRL)}, vol.~36, no.~6, pp. 829--849, 1989.

\bibitem{Hui&others2003}
K.-P. Hui, N.~Bean, M.~Kraetzl, and D.~Kroese, ``Network reliability estimation
  using the tree cut and merge algorithm with importance sampling,'' in
  \emph{Proceedings of the 4th International Workshop on Design of Reliable
  Communication Networks}.\hskip 1em plus 0.5em minus 0.4em\relax Banff,
  Alberta, Canada: IEEE, 2003, pp. 254--262.

\bibitem{Papaioannou2021}
I.~Papaioannou, ``Lecture notes in {A}dvanced {S}tochastic {F}inite {E}lement
  {M}ethods,'' Munich, Germany, 2021.

\bibitem{Chen&Liu1997}
S.~X. Chen and J.~S. Liu, ``Statistical applications of the {P}oisson-binomial
  and conditional {B}ernoulli distributions,'' \emph{Statistica Sinica},
  vol.~7, no.~4, pp. 875--892, 1997.

\bibitem{Chen&others1994}
X.-H. Chen, A.~P. Dempster, and J.~S. Liu, ``Weighted finite population
  sampling to maximize entropy,'' \emph{Biometrika}, vol.~81, no.~3, pp.
  457--469, 1994.

\bibitem{Gail&others1981}
M.~H. Gail, J.~H. Lubin, and L.~V. Rubinstein, ``Likelihood calculations for
  matched case-control studies and survival studies with tied death times,''
  \emph{Biometrika}, vol.~68, no.~3, pp. 703--707, 1981.

\bibitem{Jasmon&Kai1985}
G.~B. Jasmon and O.~S. Kai, ``A new technique in minimal path and cutset
  evaluation,'' \emph{IEEE Transactions on Reliability}, vol. R-34, no.~2, pp.
  136--143, 1985.

\bibitem{Ford&Fulkerson1956}
L.~R. Ford and D.~R. Fulkerson, ``Maximal flow through a network,''
  \emph{Canadian Journal of Mathematics}, vol.~8, pp. 399--404, 1956.

\bibitem{Stoer&Wagner1997}
M.~Stoer and F.~Wagner, ``A simple min-cut algorithm,'' \emph{Journal of the
  ACM (JACM)}, vol.~44, no.~4, pp. 585--591, 1997.

\bibitem{Burer&Letchford2012}
S.~Burer and A.~N. Letchford, ``Non-convex mixed-integer nonlinear programming:
  {A} survey,'' \emph{Surveys in Operations Research and Management Science},
  vol.~17, no.~2, pp. 97--106, 2012.

\bibitem{Mitchell1998}
M.~Mitchell, \emph{An Introduction to Genetic Algorithms}.\hskip 1em plus 0.5em
  minus 0.4em\relax MIT press, 1998.

\bibitem{Yeh2008}
W.-C. Yeh, ``A fast algorithm for searching all multi-state minimal cuts,''
  \emph{IEEE Transactions on Reliability}, vol.~57, no.~4, pp. 581--588, 2008.

\bibitem{LEcuyer1994}
P.~L'Ecuyer, ``Efficiency improvement and variance reduction,'' in
  \emph{Proceedings of Winter Simulation Conference}.\hskip 1em plus 0.5em
  minus 0.4em\relax San Diego, CA, USA: IEEE, 1994, pp. 122--132.

\bibitem{Grainger1999}
J.~J. Grainger, \emph{Power System Analysis}.\hskip 1em plus 0.5em minus
  0.4em\relax McGraw-Hill, 1999.

\bibitem{Crucitti&others2004}
P.~Crucitti, V.~Latora, and M.~Marchiori, ``Model for cascading failures in
  complex networks,'' \emph{Physical Review E}, vol.~69, no.~4, p. 045104,
  2004.

\bibitem{Miao2018}
H.~Miao, ``The seismic response and dynamic function reliability analysis of
  underground water supply networks (in {C}hinese),'' PhD thesis, Tongji
  University, 2018.

\bibitem{Wang1991}
D.~Wang, ``A preliminary research on the damage prediction of the buried line
  (in {C}hinese),'' \emph{Journal of Zhengzhou Institute of Technology},
  vol.~12, no.~1, pp. 65--68, 1991.

\end{thebibliography}
%% requested to list  a bibtex style file in the manuscript if they do





% \end{thebibliography}

\end{document}